\documentclass{jfm}
\usepackage{graphicx}
\usepackage{epstopdf, epsfig}

\usepackage{amsmath, wasysym, verbatim, bbm, color, graphics, physics, bm, bbold, mathtools}
\usepackage{braket}
\usepackage[table]{xcolor}

\usepackage{csquotes}

\usepackage{booktabs}
\usepackage{array}
\usepackage{makecell}
\usepackage{multirow}
\newcolumntype{L}[1]{>{\raggedright\let\newline\\\arraybackslash\hspace{0pt}}m{#1}}
\newcolumntype{C}[1]{>{\centering\let\newline\\\arraybackslash\hspace{0pt}}m{#1}}
\newcolumntype{R}[1]{>{\raggedleft\let\newline\\\arraybackslash\hspace{0pt}}m{#1}}

\usepackage{hyperref}
\usepackage{bookmark}
\hypersetup{
    colorlinks=true,
    citecolor=blue,      
    linkcolor=blue,
}

\title{Trading particle shape with fluid symmetry: \\ on the mobility matrix in 3D chiral fluids}

\author{Tali Khain\aff{1,2},
    Michel Fruchart\aff{1,2,3},
    Colin Scheibner\aff{1,2,4,5},
    Thomas A. Witten\aff{1,2},
    \and Vincenzo Vitelli\aff{1,2,6}\footnote{Email address for correspondence: vitelli@uchicago.edu}}
\affiliation{
\aff{1}James Franck Institute, The University of Chicago, Chicago, IL 60637, USA
\aff{2}Department of Physics, The University of Chicago, Chicago, IL 60637, USA
\aff{3}Gulliver, ESPCI Paris, Université PSL, CNRS, 75005 Paris, France
\aff{4}Center for the Physics of Biological Function, Princeton University, Princeton, NJ 08544, USA
\aff{5}Princeton Center for Theoretical Science, Princeton University, Princeton, NJ 08544, USA
\aff{6}Kadanoff Center for Theoretical Physics, The University of Chicago, Chicago, IL 60637, USA
}

\begin{document}

\maketitle

\begin{abstract}
Chiral fluids - such as fluids under rotation or a magnetic field as well as synthetic and biological active fluids - flow in a different way than ordinary ones. Due to symmetries broken at the microscopic level, chiral fluids may have asymmetric stress and viscosity tensors, for example giving rise to a hydrostatic torque or non-dissipative (odd) and parity-violating viscosities. 
In this article, we investigate the motion of rigid bodies in such an anisotropic fluid in the incompressible Stokes regime through the mobility matrix, which encodes the response of a solid body to forces and torques. 
We demonstrate how the form of the mobility matrix, which is usually determined by particle geometry, can be analogously controlled by the symmetries of the fluid.
By computing the mobility matrix for simple shapes in a three-dimensional anisotropic chiral fluid, we predict counter-intuitive phenomena such as motion at an angle to the direction of applied forces and spinning under the force of gravity.
\end{abstract}

\section{Introduction}

The motion of small particles in a fluid, from many-body sedimentation~\citep{Ramaswamy2001, Guazzelli2009, Goldfriend2017screening, Chajwa2019kepler} to the swimming of micro-organisms~\citep{Purcell1977life,Taylor1951, lauga2009hydrodynamics, lauga2020fluid}, is described by the physics of low Reynolds number flows. 
In this regime, viscous forces dominate over inertial forces, which simplifies the full Navier-Stokes equation to the linear Stokes equation,
\begin{align}
    0 &= F^{\text{ext}}_i + \partial_j \sigma_{ij} \label{eq:Stokes}\\
    \sigma_{ij} &= \sigma_{ij}^\text{h} + \eta_{ijk\ell} \partial_\ell v_k
    \label{eq:stress}
\end{align}
Here, $\sigma_{ij}$ is the stress tensor, which consists of two contributions: a hydrostatic stress $\sigma_{ij}^\text{h}$, which can be non-zero even in an undisturbed or uniform flow, and a viscous stress, which arises in response to gradients in velocity.
In a standard fluid, such as air or water, the hydrostatic stress is due to the pressure, $\sigma_{ij}^\text{h} = -P \delta_{ij}$, and the tensorial viscosity $\eta_{ijk\ell}$ reduces to just a few coefficients, such as the bulk and shear viscosities.

\begin{figure}
    \centering
    \includegraphics[width=0.9\columnwidth]{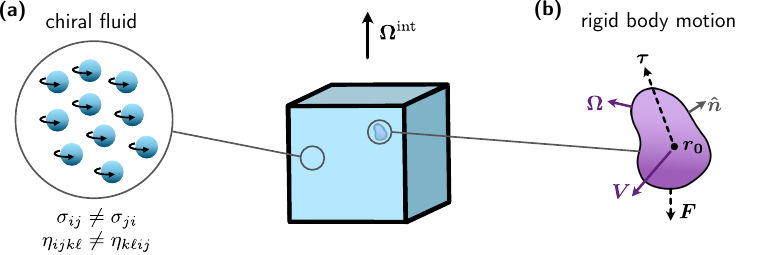}
    \caption{\label{fig:setup}
    \textbf{Rigid body motion in a chiral fluid.} (a) A fluid composed of particles driven to spin with angular velocity $\bm{\Omega^{\text{int}}}$, for example due to an external magnetic field, is chiral. Since the microscopic constituents of the fluid break parity or mirror symmetry, the stress and viscosity tensors of the fluid no longer need be symmetric. (b) As a result, the way a rigid body moves and rotates under applied forces and torques is modified.}
\end{figure}

Not all fluids share this property. Depending on the symmetries obeyed by the microscopic constituents of the fluid, its stress tensor (Eq.~\ref{eq:stress}) could contain additional terms. In this article, we consider chiral fluids, or fluids that break parity or mirror symmetry at the microscopic level (Fig.~\ref{fig:setup}a). Examples include polyatomic gases under magnetic fields~\citep{Korving1967influence}, magnetized plasma~\citep{Chapman1939}, fluids under rotation~\citep{Yoshinari1956Kinetic}, vortex~\citep{wiegmann2014anomalous} and electron fluids~\citep{Berdyugin2019measuring, Bandurin2016}, ferrofluids~\citep{Reynolds2023}, and active and driven systems made up of spinning particles~\citep{Condiff1964, Tsai2005chiral, soni2019odd, reeves2021emergence,Hargus2020}.

In these chiral systems, the stress and the viscosity tensors are no longer constrained to be symmetric,
\begin{align}
    \sigma_{ij} & \neq \sigma_{ji}\\
    \eta_{ijk\ell} & \neq \eta_{k\ell ij}
\end{align}
As a result, both quantities $\sigma^{\text{h}}_{ij}$ and $\eta_{ijk\ell}$ can acquire additional contributions. The hydrostatic stress $\sigma^{\text{h}}_{ij}$ can contain an antisymmetric term of the form $-\epsilon_{ijk}\chi_k$, where $\epsilon_{ijk}$ is the Levi-Civita tensor, and $\chi_k$ corresponds to an ambient torque density, which can arise, for example, in a fluid of spinning particles. Moreover, the non-zero anti-symmetric part of the viscosity tensor, referred to as the odd viscosity, generates a viscous response which is non-dissipative~\citep{avron1998odd,Lapa2014swimming, ganeshan2017odd, markovich2022non,fruchart2023odd}.

Such a relaxation of the symmetry constraints leads to rich flow behavior in three dimensions at low Reynolds number~\citep{khain2022stokes}. 
The presence of the torque density and odd viscosity alters how the chiral fluid responds to applied deformations, such as when flowing past a rigid body. Consequently, the motion of the body itself through a chiral fluid changes as well.
In the overdamped regime of Stokes flow, the motion of a solid body in a fluid is described by its linear and angular velocities $\bm{V}$ and $\bm{\Omega}$~\citep{Happel1983low, KimKarrila} (Fig.~\ref{fig:setup}b). Given a reference point on the object $\bm{r_0}$, the velocity at any other point on the object is $\bm{v}(\bm{r}) = \bm{V} + \bm{\Omega} \times (\bm{r} - \bm{r_0})$. Due to the linearity of the Stokes equations, the forces $\bm{F}$ and torques $\bm{\tau}$ on the solid body as it moves are linear functions of $\bm{V}$ and $\bm{\Omega}$. The coefficient of proportionality is the resistance matrix $\mathbb{P}$ (also called the propulsion matrix), such that $\mathcal{F} = \mathbb{P} \mathcal{V}$, where $\mathcal{F} = [\bm{F}, \bm{\tau}]^T$ and $\mathcal{V} = [\bm{V}, \bm{\Omega}]^T$. The inverse of $\mathbb{P}$, called the mobility matrix, $\mathbb{M}$, gives us the response of the solid body to external forces and torques\footnote{The external forces and torques on the solid body ($\bm{F}, \bm{\tau}$) are balanced by the hydrodynamic forces (the forces on the solid body by the fluid), which we denote by $\bm{F}^{\text{fluid}}, \bm{\tau}^{\text{fluid}}$. In our notation, then, $\mathcal{V} = \mathbb{M} \mathcal{F} = - \mathbb{M} \mathcal{F}^{\text{fluid}}$.}, 
\begin{equation}
    \begin{bmatrix}
    \bm{V}\\
    \bm{\Omega}
    \end{bmatrix}
    = 
    \mathbb{M}
    \begin{bmatrix}
    \bm{F}\\
    \bm{\tau}
    \end{bmatrix}.
    \label{eq:mobility}
\end{equation}
Here and throughout most of what follows, we assume that the fluid is quiescent in the absence of the immersed object.
(Otherwise, in Eq.~\ref{eq:mobility}, the velocity and angular velocity of the object would be with respect to the background flow of the fluid.)
The mobility matrix $\mathbb{M}$ is a $6 \times 6$ matrix, which can be split into four blocks:
\begin{equation}
    \mathbb{M}
    = 
    \begin{bmatrix}
    \begin{array}{@{}c|c@{}}
    A & B \\
    \hline
    T & S
    \end{array}
    \end{bmatrix}.
    \label{eq:mobility_matrix}
\end{equation}

Here, $\mathbb{M}$ depends both on the geometry of the object and the properties of the fluid, as well as the choice of reference point.
In order to explain particle motion, we often turn to its geometry \citep{guyon2015physical}. 
In water, for example, applying a force to the center of a sphere does not cause it to rotate; mathematically, this is captured by the fact that $T=B=0$ when the reference point is taken to be the center of the sphere.
To couple translational and rotational modes, one needs to consider more complex geometries of the solid body~\citep{brenner1965coupling, makino2003sedimentation, krapf2009chiral, palusa2018sedimentation} or go beyond hydrodynamic interactions~\citep{morozov2017dynamics}.

Yet, the geometry of the object does not fully control its mobility: the symmetries of the fluid can play a significant role.
This naturally raises the question: can we manipulate solid body motion by designing the fluid, rather than by changing the particle shape?
To address this question, we must consider the symmetries of both the fluid and the solid body, and ask how these constrain the form of the mobility matrix.

In this article, we demonstrate that one can achieve complex modes of motion of simple objects by immersing them in a chiral fluid. 
Following the setup in \citet{khain2022stokes}, we consider a three-dimensional fluid with cylindrical symmetry about the $z$-axis and standard shear viscosity $\mu$, and focus on the effect of an odd shear viscosity denoted by $\eta^{\text{o}}$.
Besides its non-dissipative nature, the coefficient $\eta^{\text{o}}$ is anisotropic (the $z$-axis is singled out), and parity-violating or chiral (when reflected across vertical planes, the coefficient acquires a minus sign). 
Each of these broken symmetries has consequences on the possible form of the mobility matrix, which we delineate in Section~\ref{sec:M_properties}.
In Section~\ref{sec:compute_M}, we review different methods of computing the mobility matrix for an arbitrary solid body for a general fluid with any viscosity $\eta_{ijk\ell}$.
We describe the simplest signature of a chiral fluid in Section~\ref{sec:odd_stress}: a chiral particle propelling under an odd hydrostatic stress (torque).
In Section~\ref{sec:lift}, we demonstrate the consequences of the sphere's asymmetric mobility matrix in the presence of odd viscosity, which includes the appearance of a lift force that generates spiraling trajectories.
In Section~\ref{sec:spinning}, we illustrate the rotation-translation coupling effect of $\eta^{\text{o}}$ with the example of a sedimenting triangle spinning under the force of gravity.
In these sections, the parity-violating and odd viscosity $\eta^{\text{o}}$ produces effects that are generated by particle geometry in a standard fluid.
We further emphasize the analogous roles of particle shape and fluid symmetries in Section~\ref{sec:trading} by providing examples of pairs of distinct systems with similar particle motion.

\section{Properties of the mobility matrix}
\label{sec:M_properties}

\subsection{Constraints from spatial symmetries}
\label{sec:M_symmetries}

In a standard isotropic fluid, such as water, the mobility matrix of an object is constrained by the object's spatial symmetries. 
A cone, for example, is unchanged under rotations about its axis, and so its mobility matrix must remain invariant under this transformation as well.
If the fluid is anisotropic, however, it is not sufficient to consider just the geometry of the object.
To determine how $\mathbb{M}$ should transform under a given coordinate transformation (such as a rotation or reflection), we must consider the effect of the transformation on both the body and the fluid\footnote{Here, we consider the ``effective" symmetry group of the fluid by looking at the symmetries obeyed by its viscosity tensor. Importantly, a cylindrically symmetric viscosity tensor is invariant under the reflection $(x, y, z) \to (x, y, -z)$, even if the constituents of the fluid are not~\citep{khain2022stokes}. Similarly, the symmetry of the mobility matrix can be higher than the symmetry of the object.}.
Only if the full system remains invariant -- if the transformation is a symmetry of both the fluid and the object -- so should $\mathbb{M}$.

In standard isotropic fluids, this subtlety does not arise, since the fluid is invariant under all rotations and reflections. In this case, it appears as if the solid body transforms ``independently" \citep{guyon2015physical}.
An analogous example occurs for a sphere in an anisotropic fluid: here, the sphere is unchanged under all distance-preserving coordinate transformations, so we need to consider only the symmetries of the fluid.

\begin{figure}
    \centering
    \includegraphics[width=\columnwidth]{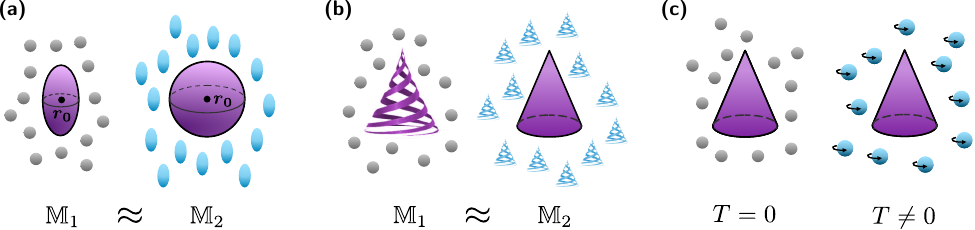}
    \caption{\label{fig:symmetries}
    \textbf{Spatial symmetries constrain the mobility matrix.} (a) An ellipsoid in an isotropic fluid (left) has an anisotropic mobility matrix, with different drag coefficients along and perpendicular to its long axis. A sphere in an anisotropic fluid composed of ellipsoids (right) has the same spatial symmetries. As a result, the two mobility matrices must have the same form. 
    (b) A conical helix in an isotropic fluid (left) has the same spatial symmetries as a cone in a fluid composed of conical helices (right). As a result, the two mobility matrices must have the same form; in particular, both objects spin under applied forces. (Note that an ellipsoid in a fluid composed of conical helices would not break sufficient symmetries, see footnote above.)
    (c) A non-chiral object (left), such as a cone, has zero $B$ and $T$ blocks in an isotropic fluid. In a parity-violating (chiral) fluid (right), the same cone now has $T \neq 0$, meaning that applied forces lead it to rotate. }
\end{figure}

To illustrate the complementary roles of the fluid and the solid body, let us consider a cylindrically symmetric ellipsoid (a prolate spheroid, left panel of Fig.~\ref{fig:symmetries}a), centered on the origin and oriented upright along the $z$-axis, immersed in an isotropic fluid. 
Since the combined fluid and object system is invariant under rotations about the $z$-axis as well as reflections across the $x$-$y$, $y$-$z$, and $x$-$z$ planes, the mobility matrix must be as well.
To satisfy these requirements, $\mathbb{M}$ must take the form (see Appendix~\ref{app:symm}):
\begin{align}
A = 
\begin{bmatrix}
A_{11} & & \\
& A_{11} & \\
& & A_{33}
\end{bmatrix}, \
B = 0, \ T = 0, \
S = 
\begin{bmatrix}
S_{11} & & \\
& S_{11} & \\
& & S_{33}
\end{bmatrix}.
\label{eq:M_anisotropic}
\end{align}
Now, there could be other fluid/object systems with these same symmetries, and hence, with the same form of $\mathbb{M}$; for example, a sphere immersed in an anisotropic fluid composed of aligned cylindrically symmetric ellipsoids (Fig.~\ref{fig:symmetries}a right). Even though the two solid bodies are different in these two cases, the way they move under applied forces and torques is qualitatively the same.
The situation is similar in Fig.~\ref{fig:symmetries}b: the conical helix in an isotropic fluid (left) and the cone in a fluid of conical helices (right) are both invariant only under cylindrical symmetry; as a result, the form of their mobility matrices must be the same.

In addition to the fluid and solid body geometry, the mobility matrix depends on the choice of reference point (see Appendix~\ref{app:ref_point} for a derivation).
Throughout this section, the transformations we consider are applied about the reference point (that is, the reference point remains invariant).
A proper choice of reference point is important to maximize the power of the symmetry analysis: for example, the mobility matrix of a sphere with the reference point at its center is isotropic, while the mobility matrix of a sphere with the reference point on its surface is not.

Additionally, note that a non-zero entry in the mobility matrix that is allowed by symmetry could, in principle, be arbitrarily small for a given solid body in a fluid.

\subsection{A chiral object vs. a chiral fluid}
\label{sec:M_parity}

In an isotropic fluid, the structure of the mobility matrix can be used to deduce the geometric properties of the solid body. 
For instance, the off-diagonal blocks $B$ and $T$ are often considered signatures of the chirality of the solid body, as they couple rotation with translation \citep{witten2020review}. 
An object is chiral if its spatially inverted image cannot be rotated back to the original. 
The mobility matrix of a non-chiral object, then, can be spatially inverted and then rotated by a suitable rotation $R$ such that it remains invariant.
Under these transformations, the mobility matrix transforms as
\begin{align}
    \begin{bmatrix}
    \begin{array}{@{}c|c@{}}
    A & B \\
    \hline
    T & S
    \end{array}
    \end{bmatrix}
    \to 
    \begin{bmatrix}
    \begin{array}{@{}c|c@{}}
    RAR^{-1} & -RBR^{-1} \\
    \hline
    -RTR^{-1} & RSR^{-1}
    \end{array}
    \end{bmatrix}.
\end{align}
Requiring $T = -RTR^{-1}$ (and the same for $B$) implies $\text{det}(T) = \text{tr}(T) = 0$, given a proper choice of reference point about which the transformations are performed.
As a result, it is possible to find asymmetric non-chiral objects that have a nonzero $T$ block that cannot be removed even by moving the reference point, such as ``taco"-like shapes with just two symmetry planes \citep{miara2024dynamics}.
However, sufficiently symmetric non-chiral objects do have a vanishing $T$ block.
For example, a cone with a reference point along the symmetry axis has $T = 0$ (Fig.~\ref{fig:symmetries}c left, see Appendix~\ref{app:symm} for more details.)

In an anisotropic fluid, even objects with a high degree of symmetry can exhibit chiral trajectories.
In this work, we focus on a cylindrically symmetric parity-violating (or chiral) fluid that breaks mirror symmetry across planes containing the $z$-axis, such as the one represented in Fig.~\ref{fig:setup}a, following the setup in \citet{khain2022stokes}.
Let us immerse a non-chiral object in this chiral fluid -- say, a cone, with its symmetry axis aligned with $z$ -- and consider the most general form of the force-rotation block $T$ of its mobility matrix.
If we choose the reference point along the cone's symmetry axis, the fluid/object system is invariant only under rotations about the $z$-axis, which constrains $T$ to take the form\footnote{A quick way to check that the form of $T$ in Eq.~\ref{eq:nonzeroT} is in fact invariant under rotations $R$ about the $z$-axis is to notice that the $x$-$y$ block of $T$ is a linear combination of the identity and the Levi-Civita matrix, as is the $x$-$y$ block of the rotation matrix $R$. Since the identity and Levi-Civita matrix each commute with themselves as well as with each other, $R$ and $T$ must commute, and so indeed $R T R^{-1} = T$.}
\begin{align}
T =
\begin{bmatrix}
T_{11} & T_{12} & 0\\
-T_{12} & T_{11} & 0\\
0 & 0 & T_{33}
\label{eq:nonzeroT}
\end{bmatrix}.
\end{align}
The nonzero entries in Eq.~\ref{eq:nonzeroT} cannot be removed even with a different choice of reference point (see Appendix~\ref{app:symm} for more details). 
As a consequence, the cone may spin under an applied force even if the torque is zero; in this case, a non-zero $T$ is a signature of the chirality of the fluid, and not of the object (Fig.~\ref{fig:symmetries}c right). 

\subsection{Positivity}
\label{sec:M_positive}

In a fluid with positive dissipative (even) viscosities, the mobility matrix $\mathbb{M}$ is positive definite, irrespective of the presence of odd viscosities. To see this, we consider the rate of energy dissipation in a fluid,
\begin{align}
\dot{W} = \int_{\mathcal{V}} (\partial_j v_i) \sigma_{ij} dV = \int_{\mathcal{V}} (\partial_j v_i) (\eta_{ijk\ell}\partial_\ell v_k) dV = \int_{\mathcal{V}} \eta_{ijk\ell}^{\text{e}} (\partial_j v_i) (\partial_\ell v_k) dV \ge 0,
\label{eq:dissipation}
\end{align}
where $\mathcal{V}$ is the fluid volume.
Only the even part of the viscosity tensor contributes to $\dot{W}$ \citep{fruchart2023odd}; the odd viscosity does not dissipate.
The rate of energy dissipation is nonnegative ($\dot{W} \geq 0$) provided that the even viscosities are positive (technically, this means that the tensor $\eta_{ijk\ell}^{\text{e}}$ acts as a positive semidefinite linear map on the space of rank two tensors \citep{khain2022stokes}).
In an anisotropic fluid, the rates of energy dissipation along different axes may not be the same.
In principle, there could exist deformation rates that do not dissipate energy.
These inviscid deformations correspond to the basis vectors of the null space of $\eta^{\text{e}}_{ijk\ell}$.
As long as such inviscid deformations do not exist, i.e. all deformation rates correspond to a finite rate of energy dissipation, we have $\dot{W} > 0$ ($\eta_{ijk\ell}^{\text{e}}$ acts as a positive definite linear map).

Now, in the case of a solid body $S$ with surface $\partial S$ and reference point $\bm{r_0}$ immersed in an infinite volume of fluid, we can rewrite Eq.~\ref{eq:dissipation} as 
\begin{align}
\dot{W} &= \int_{\mathbb{R}^3 \backslash S} (\partial_j v_i) \sigma_{ij} dV = \int_{\mathbb{R}^3 \backslash S} \partial_j (v_i \sigma_{ij}) - v_i \partial_j \sigma_{ij} dV  = \int_{\mathbb{R}^3 \backslash S} \partial_j (v_i \sigma_{ij}) dV \\
&= \oint_{\partial S} \bm{v}\cdot \sigma \cdot (-\bm{\hat{n}})\ dA = -\oint_{\partial S} [\bm{V} + \bm{\Omega} \times (\bm{r} - \bm{r_0})] \cdot \sigma \cdot \bm{\hat{n}}\ dA\\ 
&= -(\bm{V} \cdot \bm{F}^{\text{fluid}} + \bm{\Omega} \cdot \bm{\tau}^{\text{fluid}}) \\
&=\mathcal{F}^T \mathbb{M} \mathcal{F},
\end{align}
where we have used the divergence theorem, $\nabla \cdot \sigma = 0$, and $\bm{F} = -\bm{F}^{\text{fluid}}, \bm{\tau} = - \bm{\tau}^\text{fluid}$.
(Note that when applying the divergence theorem, the outward normal is associated with the fluid volume, and points into the surface of the solid body.)
From this, we see that if there are no inviscid deformations, $\mathcal{F}^T \mathbb{M} \mathcal{F} > 0$, and so $\mathbb{M}$ is positive-definite. In the theoretical limit of only a non-dissipative viscosity, $\dot{W} = 0$ can arise, in which case $\mathbb{M}$ is only positive semi-definite.

\subsection{Symmetry}
\label{sec:M_symmetry}

We now show that $\mathbb{M} = \mathbb{M}^T$ if and only if there is no odd viscosity ($\eta_{ijk\ell}^{\text{o}} = 0$) with the help of the Lorentz reciprocal theorem \citep{Happel1983low,masoud2019reciprocal}.
In normal fluids, this theorem links two systems, where one is the ``source" (which produces a force), and the other is a ``receiver" (which measures the velocity field).
The Lorentz reciprocal theorem is satisfied only in the absence of an odd viscosity \citep{khain2022stokes}.
However, generalizations can be obtained~\citep{yuan2023stokesian, hosaka2023lorentz}, for instance by considering two systems that have odd viscosities of opposite sign.

To see this, we introduce the following energetic inner product between the velocity field and stress tensor of two systems:

\begin{align}
    \dot{W}_{1,2} &=  \int_{\mathbb{R}^3 \backslash S} \partial_j v_i^{(1)} \sigma_{ij}^{(2)} dV
    = \int_{\mathbb{R}^3 \backslash S} \partial_j v_i^{(1)} \eta_{ijk\ell}^{(2)} \partial_\ell v_k^{(2)} dV \\
    &= \int_{\mathbb{R}^3 \backslash S} \partial_j v_i^{(1)} (\eta_{ijk\ell}^{\text{e},(2)} + \eta_{ijk\ell}^{\text{o}, (2)}) \partial_\ell v_k^{(2)} dV 
\end{align}
where we have used the divergence theorem, the field equations $\nabla \cdot \sigma= 0$ and $\nabla \cdot \bm{v} = 0$, and defined $\eta_{ijk\ell}^{\text{o}} = \frac{1}{2}(\eta_{ijk\ell} - \eta_{k\ell ij})$.

Similarly,
\begin{align}
    \dot{W}_{2,1} &=  \int_{\mathbb{R}^3 \backslash S} \partial_j v_i^{(2)} \sigma_{ij}^{(1)} dV
    = \int_{\mathbb{R}^3 \backslash S} \partial_j v_i^{(2)} \eta_{ijk\ell}^{(1)} \partial_\ell v_k^{(1)} dV \\
    &= \int_{\mathbb{R}^3 \backslash S} \partial_j v_i^{(2)} (\eta_{ijk\ell}^{\text{e},(1)} + \eta_{ijk\ell}^{\text{o}, (1)}) \partial_\ell v_k^{(1)} dV \\
    &= \int_{\mathbb{R}^3 \backslash S} \partial_j v_i^{(1)} (\eta_{ijk\ell}^{\text{e},(1)} - \eta_{ijk\ell}^{\text{o}, (1)}) \partial_\ell v_k^{(2)} dV 
\end{align}
where we have interchanged $ij \leftrightarrow k\ell$ for the last equality.

Then, the difference is
\begin{align}
\dot{W}_{1,2} - \dot{W}_{2,1} = \int_{\mathbb{R}^3 \backslash S} (\eta_{ijk\ell}^{\text{e},(2)} - \eta_{ijk\ell}^{\text{e},(1)} + \eta_{ijk\ell}^{\text{o}, (2)} + \eta_{ijk\ell}^{\text{o}, (1)})\partial_j v_i^{(1)}  \partial_\ell v_k^{(2)} dV 
\label{eq:reciprocity}
\end{align}
The Lorentz reciprocal theorem states that $\dot{W}_{1,2} = \dot{W}_{2,1}$.
From Eq.~\ref{eq:reciprocity}, if the two viscosities are the same ($\eta_{ijk\ell}^{(1)} = \eta_{ijk\ell}^{(2)}$), the theorem holds only in the absence of odd viscosity.
However, if the two systems have equal even viscosities and opposite odd viscosities ($\eta_{ijk\ell}^{(1)} = \eta_{k\ell ij}^{(2)}$), such as if the external magnetic field is flipped, the theorem is satisfied.

Let us suppose that the viscosities of the two systems are equal, and choose the velocity field $\bm{v}^{(1)}$ to correspond to the velocity of a rigid body with surface $S$. 
Then, as in Sec.~\ref{sec:M_positive},
\begin{align}
    \dot{W}_{1,2} &= -\oint_{\partial S} \bm{v}^{(1)}\cdot \sigma^{(2)} \cdot \bm{\hat{n}}\ dA \\
    &= -\oint_{\partial S} [\bm{V_1} + \bm{\Omega_1} \times (\bm{r} - \bm{r_0})] \cdot \sigma^{(2)} \cdot \bm{\hat{n}}\ dA \\
    &= \bm{V_1} \cdot \bm{F_2} + \bm{\Omega_1} \cdot \bm{\tau_2} \\
    &=\mathcal{F}_2^T \mathbb{M} \mathcal{F}_1
\end{align}
From Eq.~\ref{eq:reciprocity}, in the absence of odd viscosity, we must have
\begin{align}
\mathcal{F}_2^T \mathbb{M} \mathcal{F}_1 &= \mathcal{F}_1^T \mathbb{M} \mathcal{F}_2\\
&=(\mathcal{F}_2^T \mathbb{M} \mathcal{F}_1)^T
\end{align}
which implies $\mathbb{M} = \mathbb{M}^T$.
Thus, $A = A^T, S = S^T,$ and $B = T^T$ if and only if $\eta_{ijk\ell}^{\text{o}} = 0$.

\subsection{Existence of centers}

As shown in Section \ref{sec:M_symmetry}, in the absence of odd viscosity, the mobility (and resistance) matrices must be symmetric, and so must be their diagonal blocks.
Meanwhile, the off diagonal blocks that couple force and angular velocity or torque and velocity are not constrained to be symmetric.
In standard fluids, however, it can be shown that there exists a unique choice of reference point, called the center of reaction or resistance, for which the torque-velocity block of the resistance matrix $\mathbb{P}$ is symmetric \citep{KimKarrila, Happel1983low}.
Similarly, the center of twist is a choice of reference point for which block $T$ of $\mathbb{M}$ is symmetric \citep{krapf2009chiral}.
Identifying these special points and working in their reference frame can be convenient for calculations; if $T$ is symmetric, it can be diagonalized by an orthonormal basis, in which the rotational motion of the rigid body can appear simpler.

Below, we show that a unique center of twist exists as long as the fluid has no inviscid deformations, that is, all deformation rates correspond to a finite rate of energy dissipation.
This condition is necessarily met in an isotropic fluid with the standard shear viscosity $\mu$.
In an anisotropic fluid, however, the dissipation rate may not be the same along all axes, in which case an inviscid deformation may exist.
The existence of the center of twist, then, is related to the properties of the even viscosities of the fluid, and does not directly depend on the presence of odd viscosities, as we will see.

From Eq.~\ref{eq:M_refpoint}, we have that under a change of reference point,
\begin{equation}
T^{\prime} = T^0 + S [\![\times \bm{R}]\!],
\end{equation}
or, in index notation,
\begin{equation}
T^{\prime}_{ij} = T^0_{ij} + \epsilon_{\ell kj}R_k S_{i\ell}
\end{equation}
where $T^{\prime}$ is computed with respect to $\bm{r_0^{\prime}}$ and $T^0$ is computed with respect to $\bm{r_0}$.
Here, we have defined $[\![ \bm{R} \times]\!]_{ik} = [\![\times \bm{R}]\!]_{ik} = \epsilon_{ijk} R_j$ such that $[\![ \bm{R} \times]\!] \bm{v} = \bm{R} \times \bm{v}$ and $\bm{v} [\![\times \bm{R}]\!] = \bm{v} \times \bm{R}$ for any vector $\bm{v}$, and $\bm{R} = \bm{r_0}^{\prime} - \bm{r_0}$.
Now, the anti-symmetric part of $T^{\prime}$ is zero when
\begin{align}
T_{ij}^0 - T_{ji}^0 &= -(\epsilon_{\ell kj} R_k S_{i \ell} - \epsilon_{\ell ki}R_k S_{j\ell})\\
&= \epsilon_{jk\ell} R_k S_{i\ell} - \epsilon_{ik\ell}R_k S_{j\ell}.
\end{align}
Applying $\epsilon_{mij}$ to both sides and using the identities
\begin{align}
\epsilon_{mij}\epsilon_{jk\ell} = \epsilon_{jmi}\epsilon_{jk\ell} = \delta_{mk}\delta_{i\ell} - \delta_{m\ell}\delta_{ik}\\
\epsilon_{mij}\epsilon_{ik\ell} = \epsilon_{ijm}\epsilon_{ik\ell} = \delta_{jk}\delta_{m\ell} - \delta_{j\ell}\delta_{mk},
\end{align}
we have
\begin{align}
\epsilon_{mij}(T_{ij}^0 - T_{ji}^0) &= R_{m}S_{ii} - R_{i}S_{im} - R_{j}S_{jm} + R_{m}S_{jj}\\
&= 2(R_m S_{ii} - R_i S_{im})\\
&= 2[\bm{R}\cdot (\text{tr}(S)\mathbb{I} - S)]_m.
\label{eq:twist}
\end{align}
The center of twist exists if the equation above has a solution for some $\bm{R}$. Whether this is the case depends on the properties of the matrix $\text{tr}(S)\mathbb{I} - S$.

In the absence of odd viscosity and any inviscid deformations, the block $S$ is symmetric and positive-definite, and can thus be diagonalized, in which case its three positive eigenvalues, $S_1, S_2, S_3$, are along the diagonal. In this basis,
\begin{align}
\text{tr}(S)\mathbb{I} - S = 
\begin{bmatrix}
S_2 + S_3 & 0 & 0\\
0 & S_1 + S_3 & 0 \\
0 & 0 & S_1 + S_2
\end{bmatrix}
\end{align}
The determinant of this matrix is always positive, and so $\text{tr}(S)\mathbb{I} - S$ is invertible. As a result, there is a unique solution for $\bm{R}$, and a unique center of twist.

The presence of odd viscosity does not affect the properties of $\text{tr}(S)\mathbb{I} - S$, assuming that there are no inviscid deformations (i.e. enough even viscosities are nonzero).
Then, even though $S$ is not symmetric, it is not anti-symmetric, and must be positive-definite. As a result, the eigenvalues of $S$ must have a positive real part. Whether or not $S$ is diagonalizable or has a Jordan block of size $2$ or $3$, the diagonal entries of $\text{tr}(S)\mathbb{I} - S$ must be positive, and thus $\text{det}(\text{tr}(S)\mathbb{I} - S) > 0$. Hence, a unique center of twist exists.

\section{Computing the mobility matrix}
\label{sec:compute_M}

In Section~\ref{sec:M_properties}, we constrained the form of the mobility matrix based on symmetry considerations.
Given a viscosity tensor $\eta_{ijk\ell}$, we would now like to explicitly obtain $\mathbb{M}$ for a desired shape.
To that end, we outline a few methods of solving for the mobility matrix, beginning with the classic boundary value problem approach.

\subsection{Boundary value problem}
\label{sec:M_bndryvalue}
In this method, we first solve the resistance (also called propulsion) problem \citep{KimKarrila}, and derive the mobility through matrix inversion. For clarity, let us name the blocks of the resistance matrix,
\begin{equation}
    \mathbb{P}
    = 
    \begin{bmatrix}
    \begin{array}{@{}c|c@{}}
    P_1 & P_2 \\
    \hline
    P_3 & P_4
    \end{array}
    \end{bmatrix}.
\end{equation}

Suppose we have a solid body moving through a fluid with constant velocity $\bm{V}$. To compute the velocity field outside the body, we move to a frame in which the body is held stationary, and thus $\bm{v}(r \to \infty) = -\bm{V}$. We solve the Stokes equations in Eq.~\ref{eq:Stokes} subject to incompressibility and the additional no-slip boundary condition on the surface $\partial S$ of the solid body, $\bm{v}(r \in \partial S) = 0$.\footnote{Note that other boundary
conditions could be appropriate, depending on the particular realization of an odd viscous fluid and the microscopic interactions between the
boundary and the fluid constituents.}

With $\bm{v}$ and $P$ in hand, we compute the stress in the fluid using Eq.~\ref{eq:stress}. Then, the force and torque on the solid body from the fluid are given by integrals over the object surface,
\begin{align}
\bm{F}^{\text{fluid}} &= \oint \sigma \cdot \hat{\bm{n}}\ dA \label{eq:force}\\
\bm{\tau}^{\text{fluid}} &= \oint (\bm{r} - \bm{r_0}) \times \sigma \cdot \hat{\bm{n}}\ dA
\end{align}
where $\bm{r_0}$ is a reference point and $\hat{\bm{n}}$ is the outward normal. 
Through the integrals above, we acquire the coefficients of proportionality between the force and torque and the velocity of the solid body, i.e. blocks $P_1$ and $P_3$ of the resistance matrix.

The blocks $P_2$ and $P_4$ are computed similarly. For this case, we suppose the solid body spins with angular velocity $\bm{\Omega}$. Solving the boundary value problem, computing the stress, and integrating over the object surface provides the forces and torques, which yields the remaining blocks.
Once the full resistance matrix is computed, we invert to obtain the mobility matrix.

In general, solving the Stokes equation analytically is feasible for shapes with high degree of symmetry, such as the sphere. For other geometries, a common approach is to numerically solve for the velocity field outside the object using finite element methods or other approaches.
In this work, we solve the boundary value problem analytically for the sphere in Section~\ref{sec:lift} in order to compute block $A$ of $\mathbb{M}$.
In general, to compute this block, it is necessary to solve for the full resistance matrix as outlined above, since $A = (P_1 - P_2 P_4^{-1} P_3)^{-1}$.
For a sphere in an odd viscous fluid, however, we find that $P_3 = 0$ (see Section~\ref{sec:lift}), and so we simply have $A = P_1^{-1}$ and the rotational half of the problem is not needed. 

\subsection{Boundary integral method}
\label{sec:M_bndryintegral}

Rather than solving for the velocity field in the bulk fluid outside of the solid body as in Section \ref{sec:M_bndryvalue}, the boundary integral method (also called the single-layer potential) reformulates the Stokes equations into integral form over the object's surface \citep{KimKarrila, Guazzelli2009}.
Conceptually, the boundary condition at the surface of the object can be thought to exert forces on the fluid (or vice versa), which bend the fluid flow around the solid body.
By building the object out of force singularities which are distributed on its surface, we can mimic the boundary condition and solve the mobility problem directly.

First, we obtain the velocity field due to a point force $\bm{f}(\bm{r}) = \bm{f} \delta^3(\bm{r})$ by solving the Stokes equation (\ref{eq:Stokes}),
\begin{align}
\bm{v}(\bm{r}) = \mathbb{G}(\bm{r}) \bm{f}
\end{align}
where $\bm{v}$ is termed the Stokeslet, and $\mathbb{G}$ is the Green's function or Oseen tensor. (For a description of the solution method for an arbitrary viscosity tensor, see \citealt{khain2022stokes}.)
 
In a fluid with shear viscosity $\mu$, the Green's function is given by
\begin{equation}
    \mathbb{G}^{(0)}_{ij}(\bm{r}) = \frac{1}{8\pi\mu r^3}(\delta_{ij} r^2 + r_i r_j)
    \label{eq:G_0}
\end{equation}

For the case of a rigid body, the external velocity field can then be expressed as
\begin{align}
v_i(\bm{r}) = \oint \mathbb{G}_{ij}(\bm{r'} - \bm{r})f_j(\bm{r'})dA(\bm{r'}),
\label{eq:bndry_integral}
\end{align}
where the integral is over the object's surface \citep{burgers1938second,yamakawa1970transport,pozrikidis1992boundary}
and
\begin{align}
F_j = \oint f_j(\bm{r'}) dA(\bm{r'})
\end{align}
is the external force on the sphere, or equivalently, the force the sphere exerts on the fluid.

In the case of the sphere, the force distribution $f_j$ can be guessed: due to the sphere's symmetry, the force distribution must be uniform, so $f_j(\bm{r'}) = F_j / 4\pi a^2$. 
The velocity field at $\bm{r} = 0$ (the center of the sphere) is the rigid body velocity of the sphere, $\bm{V}$, which we are looking for. 
Then, 
\begin{align}
V_i = v_i(0) = \frac{F_j}{4\pi a^2}\oint \mathbb{G}_{ij}(\bm{r'})dA(\bm{r'}).
\end{align}
From this, we read off the mobility matrix block $A$ to be
\begin{align}
A_{ij} = \frac{1}{4\pi a^2}\oint \mathbb{G}_{ij}(\bm{r'})dA(\bm{r'}).
\label{eq:A_real}
\end{align}
Computing the mobility matrix with this method requires calculating the Green's function $\mathbb{G}$ in real space, as done in \citet{yuan2023stokesian}.

An alternative approach is to carry out the integration over the sphere surface in Fourier space, which avoids the extra residue integration required to compute $\mathbb{G}$ in real space.
In Fourier space, the velocity field due to a force distribution $\bm{f}$ is given by
\begin{align}
v_i(\bm{q}) = G_{ij}(\bm{q}) f_j(\bm{q})
\end{align}
But then,
\begin{align}
v_i(\bm{r}) = \frac{1}{(2\pi)^3}\int G_{ij}(\bm{q}) f_j(\bm{q}) e^{i\bm{q}\cdot \bm{r}} d^3 \bm{q}.
\label{eq:v_integral}
\end{align}
In this case, constraining the force singularities to lie on the surface of the sphere amounts to choosing 
\begin{align}
f_j(\bm{r}) = f(r) \bm{e}_j = \frac{F_j}{4\pi a^2}\delta(r - a).
\end{align}
In Fourier space, the force is
\begin{align}
f_j(\bm{q}) = F_j \frac{\sin{(qa)}}{qa}
\end{align}
Inserting this expression into Eq.~\ref{eq:v_integral} and evaluating at $r = 0$ as before yields
\begin{align}
A_{ij} = \frac{1}{(2\pi)^3}\int G_{ij}(\bm{q}) \frac{\sin{(qa)}}{qa} d^3\bm{q}
\end{align}
Obtaining the mobility matrix in this way is referred to as the shell localization method in \citet{levine2001response,lier2023lift}.
Note that this method works only for a sphere under a force; if there is also a torque applied, it may not be sufficient to just consider the Stokeslets to compute the rigid body velocity.

\subsection{Discrete Stokeslet method}
\label{sec:M_shape}

In Sections~\ref{sec:M_bndryvalue}-\ref{sec:M_bndryintegral}, we reviewed two methods for computing the mobility matrix that can be solved analytically in the case of a sphere. 
For rigid bodies with fewer degrees of symmetry, however, analytically solving the boundary value problem in Section~\ref{sec:M_bndryvalue} is generally intractable.
Meanwhile, although the boundary integral formulation of Eq.~\ref{eq:bndry_integral} applies to asymmetric shapes, the force singularity distribution may no longer be uniform and is in general unknown.
To overcome this problem, we describe a discrete variation of the singularity method of Section~\ref{sec:M_bndryintegral} in which the forces can be solved for given a choice for the positions of the Stokeslets (or in this case, small spheres).
This method bears similarity to the immersed boundary method, in which the boundary conditions at the surface of an object are modeled through a forcing function~\citep{mittal2005immersed,Verzicco2023}.
In our case, to determine the force magnitude and direction associated with each Stokeslet, we follow the formalism originally developed for modeling polymer chains in~\citet{kirkwood1948intrinsic,bloomfield1967frictional,rotne1969variational, yamakawa1970transport, meakin1987properties}, more recently used for general rigid bodies and reviewed in~\citet{krapf2009chiral, mowitz2017predicting, witten2020review} and reproduce the method below. Our code implementing this method is available at \url{https://doi.org/10.5281/zenodo.12556863}.

As before, the goal is to compute the velocity of a rigid body moving in a fluid under an external force $\bm{F}$.
We begin by covering the object with a distribution of small spheres of radius $a$.
Assuming that the distance between the spheres is significantly larger than their size, we can neglect the near-field velocity field generated by the spheres, and treat each one as a Stokeslet under a yet undetermined force $\bm{f^{\alpha}}$.  
Given a distribution of $N$ Stokeslets, the velocity field at the position of Stokeslet $\alpha$ is given by a linear superposition of velocity fields generated by the remaining Stokeslets,
\begin{align}
    \bm{v}^{\alpha}(\bm{r}^{\alpha}) = \mathbb{G}(\bm{r}^{\alpha}-\bm{r}^{\beta})\bm{f}^{\beta}
\end{align}

where $\mathbb{G}$ is the Green's function and $\bm{f}$ is the applied force (note the summation over $\beta$).
If these were free Stokeslets, we would simply impose that Stokeslet $\alpha$ move with velocity $\bm{v}^\alpha$. However, since these Stokeslets model a rigid body, and are thus rigidly connected, Stokeslet $\alpha$ moves instead with velocity
\begin{align}
    \bm{u}^{\alpha}(\bm{r}^\alpha) = \bm{V}(\bm{r_0}) + \bm{\Omega} \times (\bm{r}^{\alpha}-\bm{r_0})
    \label{eq:app_rigidbody}
\end{align}
where $\bm{V}$ and $\bm{\Omega}$ are the velocity and angular velocity of the rigid body, respectively, and $\bm{r_0}$ is a reference point on the object.

Due to the mismatch of the velocities of the Stokeslet and the fluid around it, the Stokeslet exerts a force on the fluid given by 
\begin{align}
    \bm{f}^\alpha = P_1 (\bm{u}^\alpha - \bm{v}^\alpha)
    \label{eq:force_Stokeslet}
\end{align}
where $P_1$ is the top left block of the resistance matrix of a small sphere with radius $a$. In the case we consider below, we find that $P_3 = 0$, so we can substitute $P_1 = A^{-1}$. 
In a standard fluid with shear viscosity $\mu$, the coefficient in Eq.~\ref{eq:force_Stokeslet} is simply Stokes drag.
This block of the mobility/resistance matrix of the sphere must be computed separately in order to apply this method, for example by using one of the techniques in Sections~\ref{sec:M_bndryvalue} or~\ref{sec:M_bndryintegral}.

Combining the previous three equations, we can write the velocity of Stokeslet $\alpha$ as 
\begin{align}
    \bm{u}^\alpha(\bm{r}^\alpha) &= \bm{V}(\bm{r_0}) + \bm{\Omega} \times (\bm{r}^{\alpha}-\bm{r_0}) \\
    &= \mathbb{G}(\bm{r}^{\alpha}-\bm{r}^{\beta})\bm{f}^{\beta} + \delta^{\alpha \beta} A \bm{f}^\beta
    \label{eq:rigidsediment}
\end{align}

Meanwhile, the external force and torque on the rigid body are given by
\begin{align}
    \bm{F} &= \sum_{\alpha}\bm{f}^{\alpha}\\
    \bm{\tau} &= \sum_{\alpha} (\bm{r}^{\alpha} - \bm{r}_0) \times \bm{f}^{\alpha}
\end{align}

\begin{figure}
    \centering
    \includegraphics[width=\columnwidth]{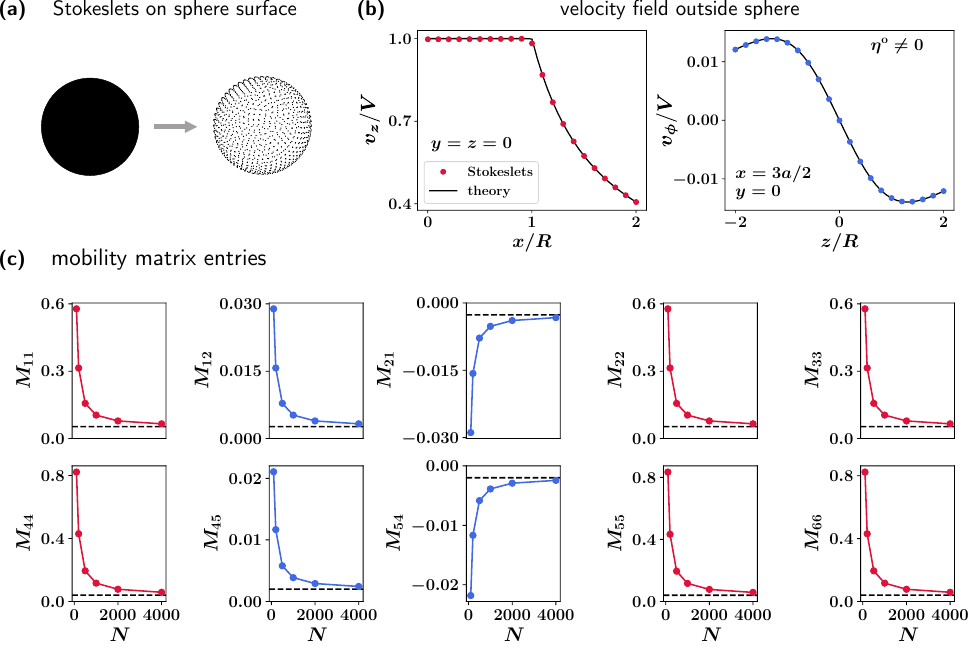}
    \caption{\label{fig:sphere_comp}
    \textbf{Comparison between theory and the Stokeslet method for modeling a rigid sphere with odd viscosity.} (a) The sphere is built out of $N$ Stokeslets uniformly distributed on its surface. (b) The flow field outside the sphere for $N=1000$ Stokeslets without (left) and with odd viscosity $\eta^{\text{o}}$ (right). By covering the sphere with ``odd" Stokeslets (Stokeslets computed in the presence of a non-zero odd viscosity), one can successfully model a rigid sphere in an odd viscous flow as done for the standard fluid. (c) The mobility matrix entries for the sphere as a function of $N$. The off-diagonal terms (in blue) are non-zero only in the presence of $\eta^{\text{o}}$. As $N$ increases, the drag coefficients converge to the theoretical values (dashed lines), even in the case of odd viscosity. For these computations, $R = 1, a = 0.001, \eta^o/\mu = 0.1$.}
\end{figure}

Since Eq.~\ref{eq:rigidsediment} is linear in the $\bm{f}$s, we can solve for $\bm{F}$ and $\bm{\tau}$ in terms of $\bm{V}$ and $\bm{\Omega}$. 
Introducing a helper $3N \times 6$ matrix $U$,
\begin{equation}
U = \begin{pmatrix}
U^1 \\
U^2 \\
\vdots \\
U^N
\end{pmatrix}
\end{equation}
with blocks
\begin{align}
    U^{\alpha} =
    \begin{bmatrix}
        1 & 0 & 0 & 0 & (z^\alpha - z_0) & -(y^\alpha - y_0)\\
        0 & 1 & 0 & -(z^\alpha - z_0) & 0 & (x^\alpha - x_0)\\
        0 & 0 & 1 & (y^\alpha - y_0) & -(x^\alpha - x_0) & 0\\
    \end{bmatrix}
\end{align}
we can write $\bm{u} = U \mathcal{V}$ and $\mathcal{F} = U^T \bm{f}$.
But then Eq.~\ref{eq:rigidsediment} takes the form
\begin{equation}
\mathbb{G}_{\text{full}} \bm{f} = U \mathcal{V} 
\end{equation}
where $\mathbb{G}_{\text{full}}$ is a $3N \times 3N$ matrix composed of $3 \times 3$ blocks. If $\alpha = \beta$, the block is simply $A$, and if $\alpha \neq \beta$, the block is given by $\mathbb{G}(\bm{r}^\alpha-\bm{r}^\beta)$.
From this, it follows that
\begin{equation}
\mathcal{F} = U^T \mathbb{G}_{\text{full}}^{-1} U \mathcal{V}
\end{equation}
and thus the mobility matrix for the rigid body is
\begin{equation}
\mathbb{M} = (U^T \mathbb{G}_{\text{full}}^{-1} U)^{-1}.
\label{eq:M_Stokeslets}
\end{equation}

Using Eqs.~\ref{eq:M_Stokeslets} and~\ref{eq:mobility}, we compute the rigid body $\bm{V}$ and $\bm{\Omega}$. Finally, to obtain the new positions of the Stokeslets, we use Eq.~\ref{eq:app_rigidbody} and integrate the overdamped equation
\begin{align}
    \dot{\bm{r}^\alpha} = \bm{u}^\alpha.
\end{align}

To demonstrate the validity of this method, we consider the case of a rigid sphere, first in a standard fluid with shear viscosity $\mu$. By uniformly covering a spherical shell with Stokeslets (Fig.~\ref{fig:sphere_comp}a) using a Fibonacci lattice~\citep{gonzalez2010measurement}, one can both recover the velocity field outside the sphere (Fig.~\ref{fig:sphere_comp}b, red) as well as the mobility matrix (Fig.~\ref{fig:sphere_comp}c, red) in the large $N$ limit.
In the Sections that follow, we use the methods described in Sections~\ref{sec:compute_M}-\ref{sec:M_shape} to illustrate the effects of a chiral fluid on particle motion, and confirm their validity in the presence of odd viscosity.

\section{Propulsion with odd hydrostatic stress}
\label{sec:odd_stress}

Chiral fluids are typically made of particles that all rotate in the same way. As a consequence, their hydrostatic stress 
\begin{equation}
    \sigma^{\text{h}}_{i j} = - p \delta_{i j} -\epsilon_{ijk} \chi_k
\end{equation}
typically contains an antisymmetric part $-\epsilon_{ijk}\chi_k$ capturing local torques in addition to the pressure. As these torques are present even in the unperturbed fluid, they usually are responsible for the most visible effects of the chirality.
We now demonstrate that a chiral object put in a chiral fluid can propel by converting the local torques into linear motion.

This can be understood as follows. We assume that $\chi_k$ is constant in time and uniform in space~\citep{banerjee2017odd, markovich2021odd, han2021fluctuating}. In this case, the torque density induces a net torque $\tau_k = 2\chi_k V$ on a solid body of volume $V$ immersed in the fluid. Therefore, we expect that the object will rotate: this is indeed what happens, and this effect is described by the block $S$ in the mobility matrix. In addition, when the block $B$ is non-zero, the presence of a net torque also leads to linear motion.

\begin{figure}
    \centering
    \includegraphics[width=\columnwidth]{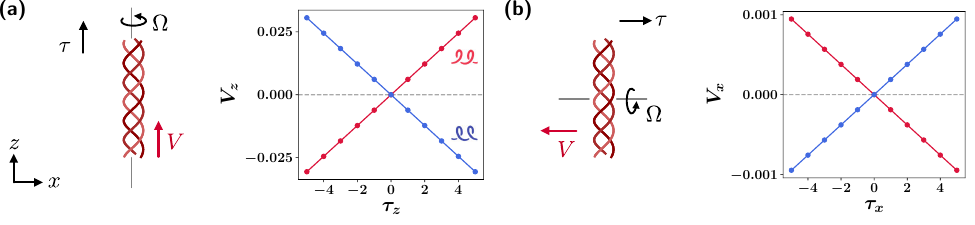}
    \caption{\label{fig:helix}
    \textbf{A triple helix propelled in a fluid with torque density.} (a) If the axis of the helix is aligned with the torque direction, the helix rotates about and moves along this axis. The velocity of propulsion is linear in $\tau$, as evident from Eq.~\ref{eq:M_helix}. (b) If the torque and helix axis are orthogonal to each other, the helix still experiences propulsion along the torque direction, although it is weaker and with an opposite sign. In both cases, the chirality of the helix (red vs. blue curves) flips the sign of the velocity response.}
\end{figure}

In a fluid with cylindrical symmetry, such as in a fluid composed of particles spinning with $\bm{\Omega^{\text{int}}} = \Omega^{\text{int}}\hat{\bm{z}}$ or under an external magnetic field in the $z$-direction, the allowed form of the hydrostatic stress is $\chi_k = \chi_z$~\citep{khain2022stokes}.
To evaluate the response of the body to this driving torque, we apply the mobility matrix formalism in Eq.~\ref{eq:mobility}. In the absence of forces, the translational velocity of the rigid object is $\bm{V} = B \bm{\tau}$, which simplifies to 
\begin{equation}
    V_i = 2V B_{iz} \chi_z 
\end{equation}
Similarly, the body's rotational velocity is $\Omega_i = 2 V S_{iz} \chi_z$. 

Consequently, any solid body with a non-zero $B$ -- such as an object with translational-rotational coupling -- spontaneously propels in a chiral fluid. Fig.~\ref{fig:helix} demonstrates the trajectory of a triple helix in a fluid with an ambient torque density. To model the helix, we cover it with a chain of Stokeslets as in Section~\ref{sec:M_shape}, where the Green's function is given by Eq.~\ref{eq:G_0}.

Each strand of the triple helix is given by the equation
\begin{align}
    x &= r \cos{(k z + \phi_0)}\\
    y &= r \sin{(k z + \phi_0)}
\end{align}
with $z \in [0, 2\pi), r = 1/2, k = 2$. The strands are rotated by $120^{\circ}$ with respect to each other by taking $\phi_0 = 0, 2\pi/3, 4\pi/3$. Each strand consists of $100$ Stokeslets of radius $a = 0.01$.
\begin{equation}
    \raisebox{-0.5\totalheight}
    {\includegraphics[width=0.9\columnwidth]{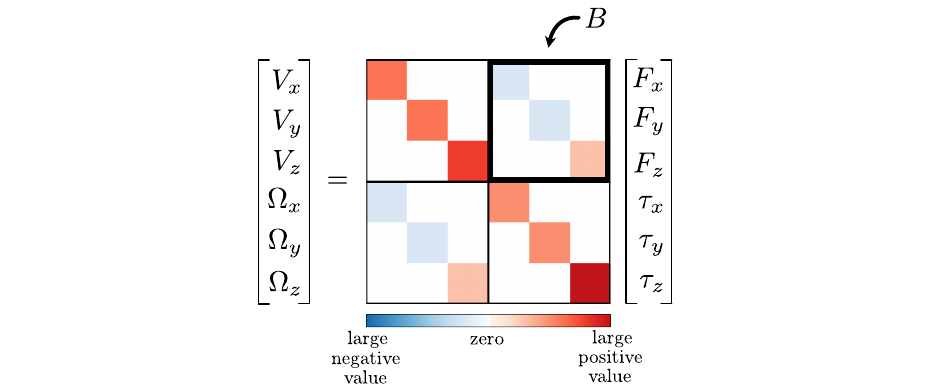}}
    \label{eq:M_helix}
\end{equation}

The mobility matrix for the helix is shown in pictorial form in Eq.~\ref{eq:M_helix}.
The colors represent the qualitative ordering of the entries, and are chosen to highlight the structure of the mobility matrix.
Red (blue) squares are positive (negative) entries, and white squares are zero. 
(For a numerical version, see Eq.~\ref{eq:app_M_helix_num} in Appendix~\ref{app:M_num}.) 
Due to nonzero entries in block $B$, the helix propels in the direction of an applied torque in addition to rotating, as shown in Fig.~\ref{fig:helix}.

The propulsion of the helix is similar to the mechanism many bacteria use to swim, which involves rotating their helical flagella \citep{lauga2016bacterial}. Here, however, the helix is driven externally by the torques in the fluid.

\section{Spiraling with a lift force}
\label{sec:lift}

In Section~\ref{sec:odd_stress}, the only role of the broken chirality in the fluid was to produce a net torque on the object through the antisymmetric part of the hydrostatic stress; for example, an immersed sphere in such a fluid would spin with constant speed \citep{khain2022stokes}.
In addition to an asymmetric stress, a chiral fluid may also have an asymmetric viscosity tensor. As shown in Section~\ref{sec:M_symmetry}, in the presence of an odd viscosity $\eta_{ijk\ell}^{\text{o}}$, the mobility matrix can also have an antisymmetric part. 

To illustrate the effect of odd viscosity and an asymmetric $\mathbb{M}$ on the motion of an immersed body, we consider the odd shear viscosities $\eta_1^{\text{o}}$ and $\eta_2^{\text{o}}$ that emerge in a three-dimensional fluid with cylindrical symmetry~\citep{khain2022stokes}, and simplify to the limit $\eta^{\text{o}} \equiv \eta_2^{\text{o}} = -\eta_1^{\text{o}}/2$
\footnote{The viscosity coefficient $\eta^{\text{o}}$ used here corresponds to the viscosity $-\frac{\bm{\ell}}{4}$ in Eq. 10 of \cite{markovich2021odd} in the limit where $\bm{\ell} = \ell_z \hat{\bm{z}}$, $-\mu_o$ in \cite{yuan2023stokesian}, and $-\frac{\eta^{\text{o}}}{2}$ in \cite{hosaka2023lorentz}.}.
In the presence of this odd viscosity, the Stokes equation takes the form 
\begin{align}
    0 = -\nabla \tilde{P} + 
    \begin{bmatrix}
    \mu && -\eta^{\text{o}} && 0 \\
    \eta^{\text{o}} && \mu && 0 \\
    0 && 0 && \mu
    \end{bmatrix}
    \Delta \bm{v}
    \label{eq:Stokes_pert}
\end{align}
where $\tilde P = P + \eta^{\text{o}} (\partial_x v_y - \partial_y v_x)$. 

To model the solid body with the Stokeslet method (Section~\ref{sec:M_shape}), we must include the effect of the odd viscosity on the velocity field.
In the limit of a small odd viscosity ($\varepsilon \equiv \eta^{\text{o}}/\mu \ll 1$), the perturbative correction to the Green's function is 
\begin{align}
    \label{eq:pert_oseen_tensor}
    \mathbb{G}^{(1)} = 
    \frac{1}{8 \pi \mu r^3}
    \begin{bmatrix}
    0 && x^2 + y^2 && yz\\
    -(x^2 + y^2) && 0 && -xz\\
    -yz && xz && 0
    \end{bmatrix}.
\end{align}

The ``odd" Stokeslet $\mathbb{G} = \mathbb{G}^{(0)} + \varepsilon \mathbb{G}^{(1)}$ has a more complex flow field than the standard point force solution. For example, under a constant force $F_z$ in the $z$-direction, the correction to the Stokeslet velocity field due to the odd viscosity is an azimuthal flow: $\bm{v} = \mathbb{G} F_z\bm{\hat{z}} = (\varepsilon F_z/8 \pi \mu) r^{-3} (yz \bm{\hat{x}} - xz \bm{\hat{y}}) = (\varepsilon F_z/8 \pi \mu) r^{-2} z \bm{\hat{\phi}}$~\citep{khain2022stokes}.

For a sphere in a standard fluid, block $A$ of the mobility matrix (top left block) is purely diagonal, giving us the familiar Stokes drag law
\begin{align}
    \bm{V} = \frac{1}{6\pi \mu a} \bm{F}.
\end{align}

To generalize this expression for a fluid with odd viscosity, we must compute the mobility matrix of the sphere using the methods in Section~\ref{sec:M_bndryvalue} or~\ref{sec:M_bndryintegral}. We begin with solving the boundary value problem (Eq.~\ref{eq:Stokes_pert}) by expanding $\bm{v} = \bm{v_0} + \varepsilon \bm{v_1}, P = P_0 + \varepsilon P_1$ (for a non-perturbative approach, see~\citet{everts2024dissipative}).
In previous work \citep{khain2022stokes}, we found that for flow in the $z$-direction, the force on the sphere is not modified to first order in $\varepsilon$. Here, we return to this problem, and compute the velocity field for the case in which the axis of the odd viscosity (taken to be $\bm{\hat{z}}$) is perpendicular to the flow. Solutions for a general $\bm{V}$ are presented below and agree with \citet{hosaka2023lorentz}\footnote{The viscosity coefficient $\eta^\text{o}$ used here corresponds to the viscosity $-\frac{\eta^{\text{o}}}{2}$ in ~\citet{hosaka2023lorentz}.}.

The known zeroth order flow past a sphere is \citep{KimKarrila}
\begin{align}
    \bm{v_0}(\bm{r}) &= -\bm{V} + 6\pi \mu a\left(1 + \frac{a^2}{6} \Delta \right) \mathbb{G^{(0)}}\cdot \bm{V}\\
    P_0 (\bm{r}) &= \frac{3a\mu}{2r^3}(\bm{V}\cdot \bm{r}).
\end{align}
The first order solution in $\varepsilon$ is a linear combination of the Green's function and its Laplacian:
\begin{align}
\bm{v_1}(\bm{r}) &= 6\pi \mu a\left(1 + \frac{a^2}{6} \Delta \right) \mathbb{G}^{(0)} \cdot \left(\bm{\hat{z}} \times \frac{\bm{V}}{2} \right)
+ 6\pi \mu a \left(1 + \frac{a^2}{6} \Delta \right)\mathbb{G}^{(1)} \cdot \bm{V} \\
P_1({\bm{r}}) &= -\frac{9a\mu}{4r^3}(\bm{z}\times \bm{V})
\end{align}
To obtain the forces on the sphere due to the fluid, we compute the stress and integrate over the surface of the sphere as in Eq.~\ref{eq:force}.
For $\bm{V} = V_x \bm{\hat{x}}$, this results in the expected Stokes drag, $F_x^{\text{fluid}} = -6\pi \mu a V_x$, at zeroth order.
In the presence of odd viscosity, the stress contains three contributions:
\begin{align}
    \sigma_{ij} = -P \delta_{ij} + \mu (\partial_i v_j + \partial_j v_i) + \sigma^{\eta^{\text{o}}}_{ij}
\end{align}
where
\begin{align}
    \sigma^{\eta^{\text{o}}} = \eta^{\text{o}}
    \begin{bmatrix}
        -2(\partial_x v_y + \partial_y v_x) && 2(\partial_x v_x - \partial_y v_y) && -\partial_y v_z - \partial_z v_y \\
        2(\partial_x v_x - \partial_y v_y) && 2(\partial_x v_y + \partial_y v_x) && \partial_x v_z + \partial_z v_x\\
        -\partial_y v_z - \partial_z v_y && \partial_x v_z + \partial_z v_x && 0
    \end{bmatrix}.
\end{align}
Notice, then, the contributions to the stress at first order in $\varepsilon$: $P_1$ contributes to the first term, $\bm{v_1}$ contributes to the second, and $\bm{v_0}$ contributes to the third. 
Carrying out the integral in Eq.~\ref{eq:force} results in a first order correction to the force on the sphere. This correction, however, is not in the direction of the uniform flow, but \textit{perpendicular} to it,
\begin{align}
    F_y^{\text{fluid}} = -3\pi \varepsilon \mu a V_x = \frac{\varepsilon}{2}F_x^{\text{fluid}}
\end{align}
Permuting indices, we see that a uniform flow in the $y$-direction gives a force in the $-x$-direction. Rewriting in terms of the external forces on the sphere, the top left block of the resistance matrix for the sphere takes the form
\begin{align}
    \begin{bmatrix}
    F_x \\
    F_y \\
    F_z 
    \end{bmatrix}
    =
    6\pi \mu a
    \begin{bmatrix}
    1 && -\varepsilon/2 && 0\\
    \varepsilon/2 && 1 && 0\\
    0 && 0 && 1
    \end{bmatrix}
    \begin{bmatrix}
    V_x\\
    V_y\\
    V_z
    \end{bmatrix},
\end{align}
which recovers the result in \cite{hosaka2023lorentz}.
To compute the top left block of mobility matrix, we invert the above equation, only keeping terms of $\mathcal{O}(\varepsilon)$,
to find,
\begin{align}
    \begin{bmatrix}
    V_x \\
    V_y \\
    V_z 
    \end{bmatrix}
    =
    \frac{1}{6\pi \mu a}
    \begin{bmatrix}
    1 && \varepsilon/2 && 0\\
    -\varepsilon/2 && 1 && 0\\
    0 && 0 && 1
    \end{bmatrix}
    \begin{bmatrix}
    F_x\\
    F_y\\
    F_z
    \end{bmatrix}
    \label{eq:A_sphere_odd}
\end{align}
Alternatively, we can obtain the same result by integrating the Green's function as in Section~\ref{sec:M_bndryintegral}, as done in~\citet{yuan2023stokesian}\footnote{The viscosity coefficient $\eta^\text{o}$ used here corresponds to the viscosity $-\mu_{o}$ in~\citet{yuan2023stokesian}.}.
Notice that the mobility matrix is indeed asymmetric, even to first order in the odd viscosity.
In Appendix~\ref{app:symm}, we further show by symmetry arguments that a cylindrically symmetric viscosity can lead to a lift force for the sphere only if it is both parity-violating and non-dissipative.

\begin{figure}
    \centering
    \includegraphics[width=\columnwidth]{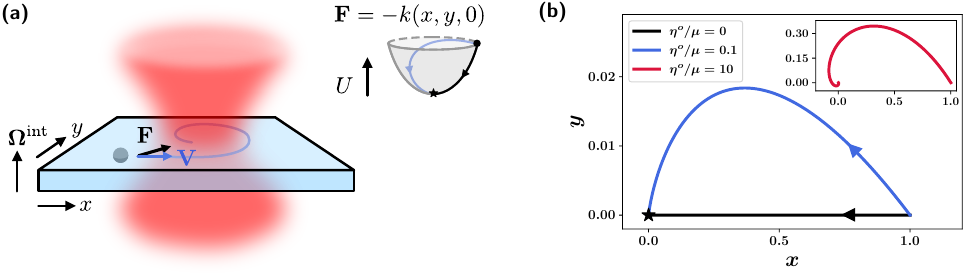}
    \caption{\label{fig:sphere}
    \textbf{A sphere experiences lift in a chiral fluid in the Stokes regime}. (a) If the direction of a force acting on the sphere is misaligned with the axis of odd viscosity ($\bm{\Omega^{\text{int}}}$), the sphere moves at angle to the force. For example, if placed in a radial potential, $U$, such as due to optical tweezers, the sphere follows a bending path to the center. 
    (b) Trajectories of the sphere from Eqs.~\ref{eq:sphere_traj_x}-\ref{eq:sphere_traj_y} with $\bm{\Omega^{\text{int}}} = \Omega^{\text{int}}_z \hat{\bm{z}}$ for different values of the odd viscosity. In the absence of odd viscosity, the sphere moves straight to the origin. If $\eta^{\text{o}} \neq 0$, the sphere's motion has a component in the $-\bm{\hat{y}}$ direction, even though the force acts in $\bm{\hat{x}}$. If the odd viscosity is significantly increased \citep{everts2024dissipative}, the sphere follows a spiraling trajectory (red).}
\end{figure}

We can now check if the method described in Section~\ref{sec:M_shape} works in the presence of odd vicosity.
To model a sphere in an odd viscous fluid, we must build it out of the modified Stokeslet in Eq.~\ref{eq:pert_oseen_tensor}, and include the modified $A$ in Eq.~\ref{eq:force_Stokeslet}. 
In Fig.~\ref{fig:sphere_comp}b (right), we demonstrate excellent agreement of the numerical and analytical velocity fields outside of the sphere.
Moreover, in Fig.~\ref{fig:sphere_comp}c, we find that the odd mobility matrix entries (in blue) computed with the Stokeslet method converge to the theoretical values at large $N$. The dashed theory lines for $\mathbb{M}_{12}$ and $\mathbb{M}_{21}$ come from Eq.~\ref{eq:A_sphere_odd}, while $\mathbb{M}_{45}$ and $\mathbb{M}_{54}$ (in block $S$) are taken from \citet{yuan2023stokesian}. 
A pictorial representation of the full mobility matrix is shown below.

\begin{equation}
    \raisebox{-0.5\totalheight}
    {\includegraphics[width=0.8\columnwidth]{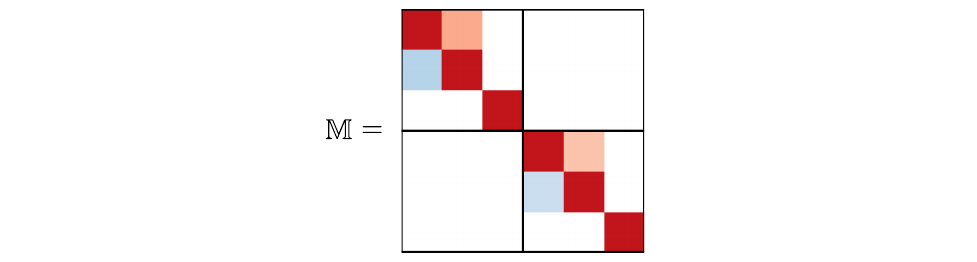}}
    \label{eq:M_sphere}
\end{equation}

Let us now examine the consequences of the anti-symmetric part of the mobility matrix of the sphere. Suppose we immerse a spherical colloidal particle in an odd viscous fluid and place it under optical tweezers (Fig.~\ref{fig:sphere}a), such that it is in a planar radial potential $\bm{F} = -k(x,y,0)$. Here, we are not interested in the $z$-motion of the sphere. If the odd viscosity axis is in the $z$-direction, we can write down the equations of motion using the mobility matrix above
\begin{align}
    \dot{x} &= -k \gamma (x + \frac{\varepsilon}{2}y)\\
    \dot{y} &= -k \gamma (-\frac{\varepsilon}{2}x + y)\\
    \dot{z} & = 0
\end{align}
where $\gamma = (6\pi\mu a)^{-1}$ and $x, y, z$ are the coordinates of the sphere.

The solutions to these equations are
\begin{align}
    x(t) &= e^{-k\gamma t}\left(x_0 \cos{(\frac{\varepsilon k\gamma t}{2})} - y_0 \sin{(\frac{\varepsilon k \gamma t}{2})}\right) \label{eq:sphere_traj_x} \\
    y(t) &= e^{-k\gamma t}\left(y_0 \cos{(\frac{\varepsilon k\gamma t}{2})} + x_0 \sin{(\frac{\varepsilon k \gamma t}{2})}\right)     \label{eq:sphere_traj_y}\\
    z(t) & = z_0
\end{align}
These trajectories are visualized in Fig.~\ref{fig:sphere}b for $x_0 = 1, y_0 = z_0 = 0$. In the absence of odd viscosity ($\varepsilon = 0$), the sphere moves to the fixed point at the origin of the potential in a straight line (black curve), as the origin is a stable node. With $\varepsilon \neq 0$, however, the lift force generated by the odd viscosity bends the trajectory (blue curve), and the fixed point becomes a stable spiral. If the odd viscosity is taken to be large, we can use the mobility matrix computed in \cite{everts2024dissipative} to show that the sphere follows a spiraling trajectory into the center of the potential (inset)
\footnote{The viscosity coefficient $\eta^{\text{o}}$ used here corresponds to the viscosity $\eta_{\text{o}}$ in \cite{everts2024dissipative} assuming that $\hat{\bm{l}} = \hat{\bm{z}}$. See Appendix \ref{app:symm} for more details.}.
Due to the anti-symmetric part of $A$, the sphere does not have three orthogonal principle axes: given the geometry in Fig.~\ref{fig:sphere}, if we apply any in-plane force, the sphere's velocity always has a component perpendicular to that force. 
(Notice that since $S$ is likewise asymmetric, the same holds for the angular velocity and torque.)
As such, these spiraling dynamics are a signature of odd viscosity. 

\section{Rotation-translation coupling}
\label{sec:spinning}

In Section \ref{sec:lift}, we demonstrated the consequences of an asymmetric mobility matrix on the example of a sphere.
Yet, even though the fluid outside a sedimenting sphere swirls (Fig.~\ref{fig:sphere_comp}b (right), \citet{khain2022stokes}), 
the $B$ and $T$ blocks of the mobility matrix are zero. 
That is, the sphere does not rotate under an applied force.
As discussed in Section~\ref{sec:M_properties}, in a standard fluid, $B$ and $T$ must be zero for non-chiral objects.
In the presence of a parity-violating viscosity, however, this constraint may no longer hold.
 
In this Section, we demonstrate this effect on an equilateral ``triangle", an object consisting of three Stokeslets that cannot move with respect to each other, immersed in a fluid with the parity-violating (and odd) viscosity $\eta^{\text{o}}$. Here, we take the triangle to be upright and in the $x$-$z$ plane. We first consider the case where the hydrostatic torque density is zero, and later add in its effect. We compute the mobility matrix following the method in Section~\ref{sec:M_shape}.

\begin{figure}
    \centering
    \includegraphics[width=\columnwidth]{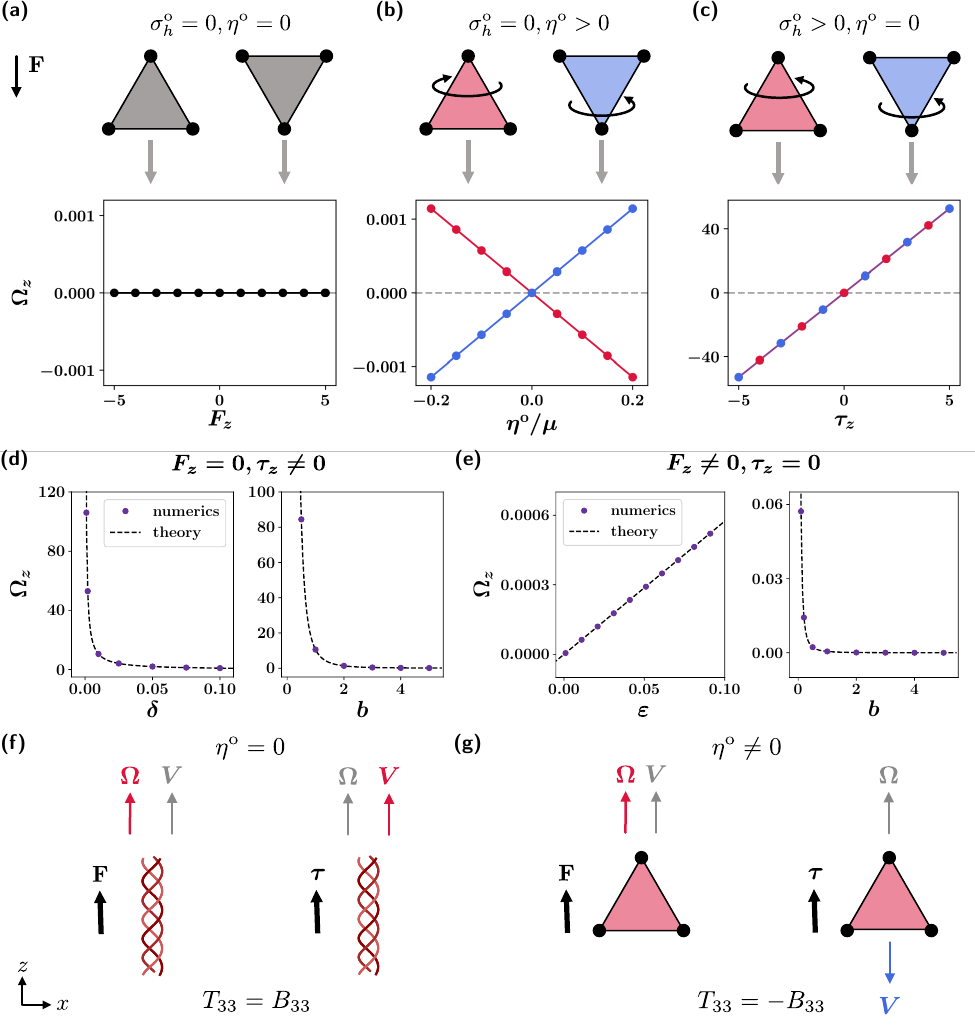}
    \caption{\label{fig:triangle}
    \textbf{A sedimenting triangle spins in a chiral fluid.}
    (a) In a standard fluid, a triangle composed of three Stokeslets, indicated by black points, falls vertically without rotating. (Note that the Stokeslet size is exaggerated for visualization purposes.)
    (b) In the presence of odd viscosity, the triangle rotates with angular speed $\Omega_z$ proportional to the applied vertical force $F_z$. The triangle rotates in the opposite direction if the odd viscosity changes sign and if the triangle's orientation is flipped. 
    (c) Under a torque density without odd viscosity, the triangle rotates in the same direction irrespective of its orientation.
    (d) The angular velocity $\Omega$ of the triangle under applied torque depends on $\delta$, the ratio of the Stokeslet radius and the side-length of the triangle $b$. In the left panel, $b = 1$, and in the right panel, $\delta = 0.01$. 
    (e) The angular velocity of the triangle under an applied vertical force depends linearly on the odd viscosity ratio $\varepsilon$ (left). Here, $b=1$ and $\delta = 0.01$. The angular velocity of the triangle likewise depends on $b$ (right). Here, $\delta = 0.01, \varepsilon  = 0.1$.
    (f) In an isotropic fluid without odd viscosity, the rotation of a helix under a force (left) is equal to the velocity of the helix under a torque (right), $T_{33} = B_{33}$. 
    (g) In a fluid with odd viscosity, the rotation of a triangle under a force is opposite to the velocity of the triangle under a torque, $T_{33} = -B_{33}$.}
\end{figure}

In the absence of $\eta^{\text{o}}$, the mobility matrix for this object is diagonal, as can be seen in Eq.~\ref{eq:M_triangle_up} (left). In Fig.~\ref{fig:triangle}a, we show that the triangle simply sinks under a vertical force, with zero angular velocity. Once $\eta^{\text{o}}$ is non-zero, however, the mobility matrix acquires several off-diagonal entries as shown in Eq.~\ref{eq:M_triangle_up} (right). (See Appendix~\ref{app:M_num} for a numerical version of $\mathbb{M}$.)
Just as for the sphere, the blocks $A$ and $S$ are asymmetric due to the non-dissipative nature of the viscosity.
Here, we focus on the newly non-zero entries in block $B$ and $T$: as a result of these, the triangle not only falls under an applied vertical force, but also rotates about the $z$-axis (Fig.~\ref{fig:triangle}b).

\begin{equation}
    \raisebox{-0.5\totalheight}
    {\includegraphics[width=0.8\columnwidth]{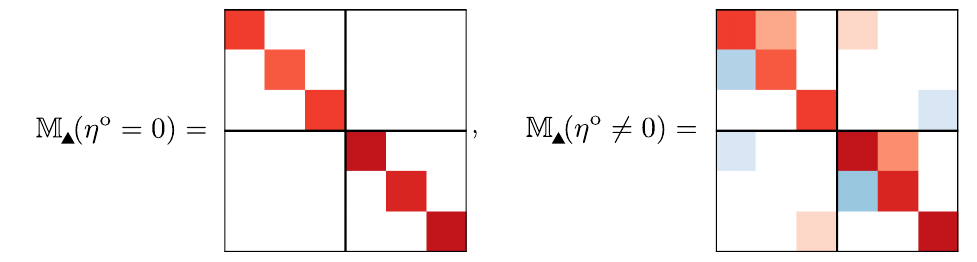}}
    \label{eq:M_triangle_up}
\end{equation}

Assuming that the distance $b$ between the Stokeslets is much larger than their characteristic size $a$, we can analytically carry out the method in Section~\ref{sec:M_shape} with the modified Green's function in Eq.~\ref{eq:pert_oseen_tensor} to compute the inverse of the mobility matrix. Expanding for small $\delta = a/b$ allow us to invert to find $\mathbb{M}$ and preserve the dependence of the mobility matrix on parameters:

\begin{small}
\begin{equation}
    \mathbb{M} = \frac{1}{b\mu \pi}
    \begin{bmatrix}
    \frac{1}{18\delta} + \frac{1}{8} - \frac{\delta}{128} & \frac{\varepsilon}{36\delta} + \frac{\varepsilon}{24} & 0 & \frac{\sqrt{3}\varepsilon}{96b} & 0 & 0 \\[3pt]
    -\frac{\varepsilon}{36\delta}-\frac{\varepsilon}{24} & \frac{1}{18\delta} + \frac{1}{12} & 0 & 0 & 0 & 0 \\[3pt]
    0 & 0 & \frac{1}{18\delta} + \frac{1}{8} - \frac{\delta}{128} & 0 & 0 & -\frac{\sqrt{3}\varepsilon}{96b} \\[3pt]
    -\frac{\sqrt{3}\varepsilon}{96b} & 0 & 0 & \frac{1}{3b^2 \delta} - \frac{1}{4b^2} & \frac{\varepsilon}{12b^2\delta} + \frac{\varepsilon}{16 b^2} & 0 \\[3pt]
    0 & 0 & 0 & -\frac{\varepsilon}{12b^2\delta}-\frac{\varepsilon}{16 b^2} & \frac{1}{6b^2 \delta} - \frac{1}{16b^2} & 0 \\[3pt]
    0 & 0 & \frac{\sqrt{3}\varepsilon}{96b} & 0 & 0 & \frac{1}{3b^2 \delta} - \frac{1}{4b^2} 
    \end{bmatrix}
    \label{eq:app_M_triangle_analyt}
\end{equation}
\end{small}
Here, we have neglected terms of order $\varepsilon^2/\delta$ and higher. For more details on the choice of $a$ and the limit of small $\delta$, see \citet{krapf2009chiral,witten2020review}.

From this mobility matrix, we read off the angular velocity of the triangle as it sediments under a vertical force:
\begin{equation}
    \Omega_z = \frac{\sqrt{3}}{96 \pi} \frac{\varepsilon}{b^2 \mu} F_z.
    \label{eq:visc_spin}
\end{equation}
Here again $\varepsilon = \eta^{\text{o}}/\mu$. 

If the chiral fluid has a torque density in addition to the parity-violating odd viscosity, as in Section~\ref{sec:odd_stress}, then the triangle rotates due to this effect as well (Fig.~\ref{fig:triangle}c). From before, $\Omega_i = 2V S_{iz} \chi_z = S_{iz} \tau_z$ where $\chi_z$ is the torque density and $\tau_z$ is the total torque. In this case, choosing $\bm{r_0}$ to be the centroid of the triangle, the angular velocity due to the torque density is
\begin{equation}
    \Omega_z = \frac{1}{3\pi b^3\mu}\left(\frac{1}{\delta} - \frac{3}{4}\right) \tau_z + \mathcal{O}\left(\frac{\varepsilon^2}{\delta}\right)
\end{equation}
where the total torque $\tau_z$ scales as $\delta^3$, the volume of the object.

The two effects can be distinguished. If the triangle is flipped upside down, the angular velocity in Eq.~\ref{eq:visc_spin} is reversed: the triangle spins in the opposite direction (Fig.~\ref{fig:triangle}b). Meanwhile, the rotation due to the torque density is independent of triangle orientation, as can be seen from the mobility matrix corresponding to the upside down triangle in Eq.~\ref{eq:M_triangle_down} and in Fig.~\ref{fig:triangle}c.
\begin{equation}
    \raisebox{-0.5\totalheight}
    {\includegraphics[width=0.8\columnwidth]{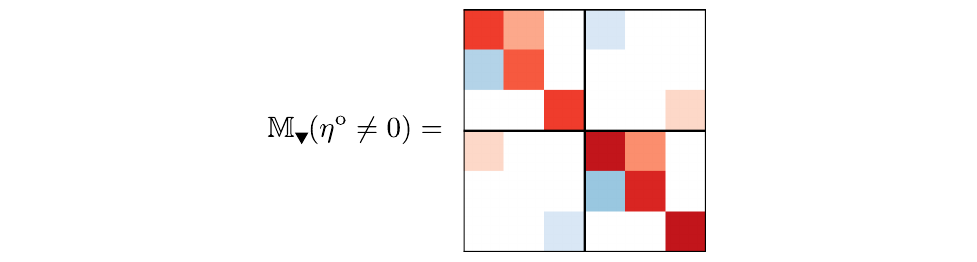}}
    \label{eq:M_triangle_down}
\end{equation}

We confirm the validity of the approximation $\delta \ll 1$ numerically by testing the dependence of two terms in the mobility matrix above on parameters $\delta, \varepsilon$, and $b$ in Fig.~\ref{fig:triangle}d, e. The dashed lines indicate the theoretical approximation from Eq.~\ref{eq:app_M_triangle_analyt}, and the markers are for the numerically computed $\mathbb{M}$.

In the above, the triangle acquires a non-zero $T$ block due to the parity-violating nature of $\eta^{\text{o}}$. 
Moreover, due to the odd nature of $\eta^{\text{o}}$, the entries in the $B$ and $T$ blocks break the symmetry of $\mathbb{M}$: as can be seen from the right of  Eq.~\ref{eq:M_triangle_up}, $B_{11} \neq T_{11}$ and $B_{33} \neq T_{33}$. 
In the absence of odd viscosity, a chiral shape can have rotation-translation coupling, but the response coefficients must be equal. For example, for the case of the helix in Eq.~\ref{eq:M_helix}, $B_{33} = V_{33}$; that is, the rotation generated by a force is equal to the velocity generated by a torque of the same magnitude (see Fig.~\ref{fig:triangle}f).
In the case of the triangle, these response coefficients are exactly antisymmetric: the velocity generated by a torque is in the opposite direction of the rotation generated by a force (Fig.~\ref{fig:triangle}g.)
In principle, for general shapes, these couplings could consist of both a symmetric and antisymmetric part, and consequently be of different magnitudes. 

\section{Trading particle shape with fluid symmetries}
\label{sec:trading}

In the previous sections, we demonstrated how complex rigid body motion can be achieved through an odd and parity-violating viscosity instead of through particle shape.
Here, we further emphasize the interplay of fluid symmetries and particle geometry by providing three concrete examples of fluid/object pairs which have the same symmetries and thus the same form of the mobility matrix.

\subsection{Anisotropic drag}
The first pair is a slender rod in an isotropic fluid and a sphere in an anisotropic fluid, where the axis of cylindrical symmetry is taken to be $z$, as in Fig.~\ref{fig:symmetries}a. (Note that a cylindrically symmetric ellipsoid, both prolate or oblate, has the same symmetries as a slender rod.)
The force on a moving rod in an isotropic fluid can be computed using slender body theory; the drag coefficient for motion along the rod's long axis is approximately twice as small as the coefficient perpendicular to it \citep{guyon2015physical}.

Similarly, in an anisotropic fluid with cylindrical symmetry, the drag coefficient of a sphere for motion along $z$ is different from the drag in the $x$ and $y$ directions.
In \citet{khain2023viscous}, we solve the boundary value problem of flow past a sphere for a fluid in which the shear viscosities $\mu_1$, $\mu_2$, and $\mu_3$ are distinct, which gives rise to anisotropic drag.

In both of these cases, the block $A$ of the mobility matrix takes the form

\begin{equation}
    \raisebox{-0.5\totalheight}
    {\includegraphics[width=0.8\columnwidth]{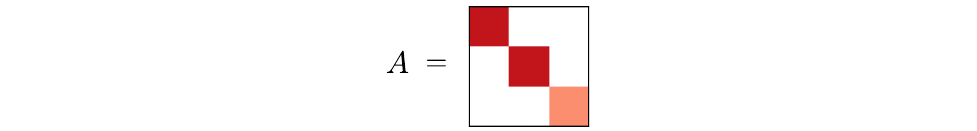}}
    \label{eq:A_aniso_drag_qual}
\end{equation}

\subsection{Rotation-translation coupling}

\begin{figure}
    \centering
    \includegraphics[width=0.6\columnwidth]{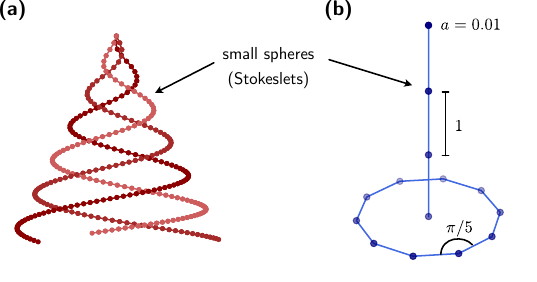}
    \caption{\label{fig:shapes_appendix}
    \textbf{Stokeslet objects}. (a) A conical helix consisting of three strands of Stokeslets. The associated mobility matrix in an isotropic fluid is given in Eq.~\ref{eq:app_M_con_helix_num}. (b) A pushpin-like object built out of Stokeslets. The associated mobility matrix in a parity-violating fluid is given in Eq.~\ref{eq:app_M_pushpin_num}.}
\end{figure}

We now consider two systems that break all spatial symmetries except cylindrical.
In Fig.~\ref{fig:symmetries}b, we argue that the mobility matrix of a conical helix in an isotropic fluid has the same form as that of a cone in a fluid composed of conical helices, namely Eq.~\ref{eq:M_cyl_even}.
We confirm this by explicitly computing the mobility matrix for two toy examples.
In the first, we construct a conical helix with three strands of $300$ equidistantly spaced Stokeslets of radius $a = 0.01$ (Fig.~\ref{fig:shapes_appendix}a), which is approximately cylindrically symmetric, and compute $\mathbb{M}$ in the presence of only the standard shear viscosity $\mu$.
The strands are given by
\begin{align}
x = c z \cos{(kz + \phi_0)}\\
y = c z \sin{(kz + \phi_0)}
\end{align}
with $c = 1/2, k = 2, z \in [0, 2\pi)$, and $\phi_0 = 0, 2\pi/3, 4\pi/3$. The reference point is along the $z$-axis.
Then, the mobility matrix takes the form in Eq.~\ref{eq:M_conical_helix_qual} (see Eq.~\ref{eq:app_M_con_helix_num} for a numerical version.)

In the second, we compute the mobility matrix for a pushpin-like object (Fig.~\ref{fig:shapes_appendix}b), which has the same symmetries as a cone (cylindrical), in a fluid with a parity-violating but even viscosity $\eta_{Q,2}^{\text{e}}$ (see \citet{khain2022stokes} for more details on this classification).
We take the viscosity $\eta_{Q,2}^{\text{e}}$ to be small as compared to the normal shear viscosity, and work at first order in $\varepsilon = \eta_{Q,2}^{\text{e}}/\mu$.

To compute the mobility matrix of the pushpin with the discrete Stokeslet method in Section~\ref{sec:M_shape}, we first compute the Green's function to first order in $\varepsilon$:
\begin{align}
\mathbb{G}^{(1)}_{11} &= \frac{x y \left(x^2+y^2-2 z^2\right)}{8 \pi \mu r^5}  \\
\mathbb{G}^{(1)}_{12} &= \mathbb{G}^{(1)}_{21} =-\frac{(x-y) (x+y) \left(x^2+y^2-2 z^2\right) }{16 \pi \mu r^5} \\
\mathbb{G}^{(1)}_{13} &= \mathbb{G}^{(1)}_{31} = -\frac{y z \left(x^2+y^2+4
   z^2\right)}{16 \pi \mu r^5}\\
\mathbb{G}^{(1)}_{22} &= \frac{-x y \left(x^2+y^2-2 z^2\right)}{8 \pi \mu r^5}\\
\mathbb{G}^{(1)}_{23} &= \mathbb{G}^{(1)}_{32} = \frac{x z \left(x^2+y^2+4 z^2\right)}{16 \pi \mu r^5} \\
\mathbb{G}^{(1)}_{33} &= 0.
\end{align}
Next, we integrate $\mathbb{G}^{(1)}$ over the surface of a sphere (Section~\ref{sec:M_bndryintegral}) to find the response of the sphere to applied forces.
We find that block $A$ of the sphere's mobility is unchanged from the isotropic case.
With these two ingredients, we compute the mobility matrix of the pushpin, which indeed takes the form in Eq.~\ref{eq:M_conical_helix_qual} (numerical values are provided in Appendix~\ref{app:M_num}).
This form can also be derived by symmetry arguments (see Eq.~\ref{eq:M_cyl_even}). 

\begin{equation}
    \raisebox{-0.5\totalheight}
    {\includegraphics[width=0.8\columnwidth]{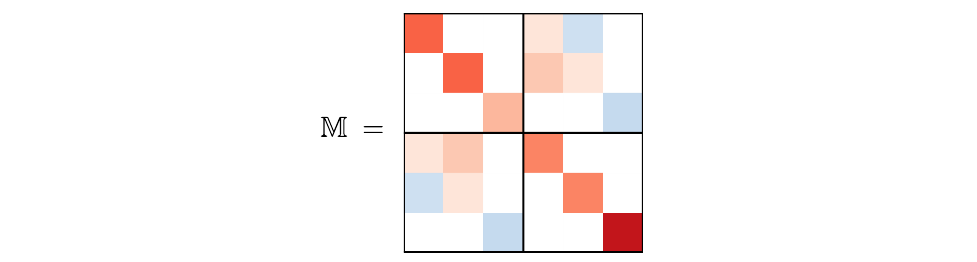}}
    \label{eq:M_conical_helix_qual}
\end{equation}

Note that in this case, the symmetry arguments hold only in the upright orientation of the conical helix and the pushpin. Once these two objects begin to rotate, the anisotropy of the fluid in the second system prevents a direct comparison of the two mobility matrices.

\subsection{Lift force}
Lastly, we consider a symmetry which is distinct from the spatial symmetries above: the symmetry of $\mathbb{M}$. 
In this case, the fluid/object pair is a sphere in a fluid with odd viscosity and a sphere under a Lorentz force in a standard isotropic fluid.

As discussed in Section~\ref{sec:M_symmetry}, the mobility matrix can acquire an anti-symmetric part in the presence of odd viscosity.
In Section~\ref{sec:lift}, we compute block $A$ of the mobility matrix for the sphere in a fluid with $\eta^{\text{o}}$, which takes the form
\begin{equation}
    \raisebox{-0.5\totalheight}
    {\includegraphics[width=0.8\columnwidth]{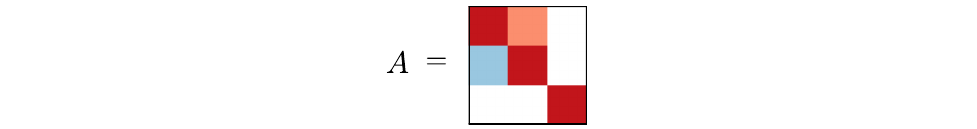}}
    \label{eq:A_odd_drag_qual}
\end{equation}

Now, suppose we immerse a solid body in a fluid with only dissipative viscosities. 
In general, one cannot break the symmetry of $\mathbb{M}$ through the geometry of the object alone; however, certain applied forces may be integrated out (i.e. solved for) in order to arrive at an effective mobility matrix $\mathbb{M_{\text{eff}}}$ that is nonsymmetric, for instance in the case of an object driven by an external field.

As an example, let us consider the overdamped motion of a charged spherical particle with charge $q$ in a quiescent fluid under a magnetic field $\bm{H}$ in the $z$-direction \citep{park2021thermodynamic}. In this case, the equation of motion for the particle is
\begin{align}
\bm{\dot{x}} = \bm{V} &= A(\bm{F} + q \bm{V}\times\bm{H})
\end{align}
where the first term on the right hand side is the mobility of an uncharged sphere, and the second term is the Lorentz force.
Rearranging, we can write
\begin{align}
\bm{V} = A_{\text{eff}} \bm{F}
\end{align}
where $A_{\text{eff}} =(\mathbb{I}+qA[\![\bm{H} \times]\!])^{-1} A$ is an effective mobility matrix induced by the external field.
If we take $A$ to be diagonal as for the sphere in an isotropic fluid and consider a small $qH_z$, we can expand $A_{\text{eff}}$ to first order in $H_z$ to find
\begin{align}
A_{\text{eff}}=
\frac{1}{6\pi\mu a}
\begin{bmatrix}
1 & \frac{qH_z}{6\pi\mu a} & 0\\
-\frac{qH_z}{6\pi\mu a} & 1 & 0\\
0 & 0 & 1 \label{eq:A_eff}
\end{bmatrix}
\end{align}
which is exactly the form of the mobility matrix in \ref{eq:A_odd_drag_qual}.

A similar asymmetric form of the mobility matrix arises for a sphere in a rotating fluid~\citep{herron1975sedimentation, tanzosh1994motion, tanzosh1995transverse}. Here, the effect of the Coriolis force causes deflection in the sphere's path which is described by the anti-symmetric off-diagonal terms, like in Eq.~\ref{eq:A_eff}.

\section{Conclusion}
In this article, we have shown that the low Reynolds number motion of rigid bodies is markedly different in a chiral fluid. The broken symmetries of the viscosity tensor in turn break symmetries of the mobility matrix, which governs particle motion. The non-dissipative nature of the viscosity allows for spiraling motion, while its parity-violation generates rotation-translation coupling even for non-chiral objects. 
We demonstrate that motion generated through particle geometry in an isotropic fluid can be mimicked by the presence of anisotropic viscosity coefficients.
By harnessing the properties of a chiral fluid, one can achieve novel methods of transport of small particles.

\medskip
\textbf{Declaration of Interests.} The authors report no conflict of interest.

\medskip
\textbf{Acknowledgements.}
We thank Howard Stone, Yael Avni, Rituparno Mandal, and Nicolas Romeo for discussions. 
T.K. acknowledges partial support from the National Science Foundation Graduate Research Fellowship under grant no. 1746045. M.F. acknowledges partial support from the National Science Foundation under grant no. DMR-2118415, a Kadanoff-Rice fellowship funded by the National Science Foundation under award no. DMR-2011854 and the Simons Foundation. V.V. acknowledges partial support from the Army Research Office under grant nos. W911NF-22-2-0109 and W911NF-23-1-0212 and the Theory in Biology program of the Chan Zuckerberg Initiative. M.F. and V.V. acknowledge partial support from the France Chicago centre through a FACCTS grant. This research was partly supported by the National Science Foundation through the Physics Frontier Center for Living Systems (grant no. 2317138) and the University of Chicago Materials Research Science and Engineering Center (award no. DMR-2011854).
\clearpage
\appendix

\section{Mobility matrix: dependence on reference point}
\label{app:ref_point}

The mobility matrix $\mathbb{M}$ depends on the choice of reference point \citep{KimKarrila}. Suppose a rigid body experiences a force $\bm{F}$ and torque $\bm{\tau}$ and moves with $\bm{V}$ and $\bm{\Omega}$ with respect to some reference point $\bm{r_0}$. If the reference point is moved to $\bm{r_0^{\prime}}$, the following relations hold:
\begin{align}
\bm{V^{\prime}} & = \bm{V} - \bm{R} \times \bm{\Omega}\\
\bm{\Omega^{\prime}} &= \bm{\Omega}\\
\bm{F^{\prime}} &= \bm{F}\\
\bm{\tau^{\prime}} &= \bm{\tau} - \bm{R} \times \bm{F}
\end{align}
where $\bm{R} = \bm{r_0}^{\prime} - \bm{r_0}$.

We can re-express these relations through block matrices,
\begin{align}
\begin{bmatrix}
\bm{F}^{\prime}\\
\bm{\tau}^{\prime}
\end{bmatrix}
&=
\begin{bmatrix}
\mathbb{I} & 0 \\
-[\![\bm{R} \times]\!] & \mathbb{I}
\end{bmatrix}
\begin{bmatrix}
\bm{F}\\
\bm{\tau}
\end{bmatrix} \label{eq:F_block}\\
\begin{bmatrix}
\bm{V}^{\prime}\\
\bm{\Omega}^{\prime}
\end{bmatrix} 
&=
\begin{bmatrix}
\mathbb{I} & -[\![\bm{R} \times]\!] \\
0 & \mathbb{I}
\end{bmatrix}
\begin{bmatrix}
\bm{V}\\
\bm{\Omega}
\end{bmatrix} \label{eq:V_block}
\end{align}
where we have defined the matrices with components $[\![ \bm{R} \times]\!]_{ik} = [\![\times \bm{R}]\!]_{ik} = \epsilon_{ijk} R_j$ such that $[\![ \bm{R} \times]\!] \bm{v} = \bm{R} \times \bm{v}$ and $\bm{v} [\![\times \bm{R}]\!] = \bm{v} \times \bm{R}$ for any vector $\bm{v}$.
Inverting Eq.~\ref{eq:F_block} and using Eq.~\ref{eq:V_block}, we can write the mobility matrix in the primed frame as
\begin{align}
\begin{bmatrix}
\bm{V}^{\prime}\\
\bm{\Omega}^{\prime}
\end{bmatrix} 
=
\begin{bmatrix}
\mathbb{I} & -[\![\bm{R} \times]\!] \\
0 & \mathbb{I}
\end{bmatrix}
\begin{bmatrix}
A & B \\
T & S
\end{bmatrix}
\begin{bmatrix}
\mathbb{I} & 0 \\
[\![\bm{R} \times]\!] & \mathbb{I}
\end{bmatrix}
\begin{bmatrix}
\bm{F}^{\prime}\\
\bm{\tau}^{\prime}
\end{bmatrix}
\end{align}
Expanding, we find that $\mathbb{M}^{\prime}$ is given by
\begin{align}
\mathbb{M}^{\prime} =
\begin{bmatrix}
A - [\![\bm{R} \times]\!]T + B[\![\times \bm{R} ]\!] - [\![\bm{R} \times]\!] S [\![\times \bm{R}]\!] & B - [\![\bm{R} \times]\!]S\\
T + S[\![\times \bm{R}]\!] & S
\end{bmatrix}
\label{eq:M_refpoint}
\end{align}
That is, all blocks depend on the choice of reference point except for the $S$ block.

\section{Mobility matrix: constraints from spatial symmetries}
\label{app:symm}

The spatial symmetries of both the fluid and solid body place constraints on the form of the mobility matrix, as discussed in Section~\ref{sec:M_symmetries}-\ref{sec:M_parity}. In this Appendix, we describe how the mobility matrix transforms under rotations and reflections, and consider the consequences of spatial symmetries in more detail.

To begin, let us suppose our system is invariant under some rotation $R$, where $R$ is a 3x3 rotation matrix. To derive how the mobility matrix transforms under $R$, we first note that the velocity and angular velocity of the solid body, as well as the applied force and torque, all transform in the same way under a rotation:
\begin{align}
\begin{bmatrix}
\bm{F}^{\prime}\\
\bm{\tau}^{\prime}
\end{bmatrix}
&=
\begin{bmatrix}
R & 0 \\
0 & R
\end{bmatrix}
\begin{bmatrix}
\bm{F}\\
\bm{\tau}
\end{bmatrix} \label{eq:F_block_R}\\
\begin{bmatrix}
\bm{V}^{\prime}\\
\bm{\Omega}^{\prime}
\end{bmatrix} 
&=
\begin{bmatrix}
R & 0 \\
0 & R
\end{bmatrix}
\begin{bmatrix}
\bm{V}\\
\bm{\Omega}
\end{bmatrix} \label{eq:V_block_R}
\end{align}
where the primes indicate the rotated frame.

Inverting Eq.~\ref{eq:F_block_R}, we can write the velocities in the primed frame as
\begin{align}
\begin{bmatrix}
\bm{V}^{\prime}\\
\bm{\Omega}^{\prime}
\end{bmatrix} 
=
\begin{bmatrix}
R & 0 \\
0 & R
\end{bmatrix}
\begin{bmatrix}
A & B \\
T & S
\end{bmatrix}
\begin{bmatrix}
R^{-1} & 0 \\
0 & R^{-1}
\end{bmatrix}
\begin{bmatrix}
\bm{F}^{\prime}\\
\bm{\tau}^{\prime}
\end{bmatrix},
\end{align}
which yields the mobility matrix
\begin{align}
\mathbb{M}^{\prime}=
\begin{bmatrix}
RAR^{-1} & RBR^{-1}\\
RTR^{-1} & RSR^{-1}
\end{bmatrix}.
\end{align}
Since the fluid/object system is invariant under $R$, the primed and unprimed mobility matrices must be the same ($\mathbb{M} = \mathbb{M}^{\prime}$), which sets constraints on the coefficients of the mobility matrix.
For example, if the fluid/object system has cylindrical symmetry, it is invariant under any rotation of the form
\begin{align}
R_z = 
\begin{bmatrix}
\cos{\phi} & \sin{\phi} & 0\\
-\sin{\phi} & \cos{\phi} & 0 \\
0 & 0 & 1
\end{bmatrix}.
\end{align}
Requiring that the mobility matrix remain unchanged under $R_z$ constrains $\mathbb{M}$ to take the form
\begin{align}
\mathbb{M}_{R_z}=
\begin{bmatrix}
A_{11} & A_{12} & 0 & B_{11} & B_{12} & 0 \\
-A_{12} & A_{11} & 0 & -B_{12} & B_{11} & 0 \\
0 & 0 & A_{33} & 0 & 0 & B_{33} \\
T_{11} & T_{12} & 0 & S_{11} & S_{12} & 0 \\
-T_{12} & T_{11} & 0 & -S_{12} & S_{11} & 0 \\
0 & 0 & T_{33} & 0 & 0 & S_{33}
\end{bmatrix}.
\label{eq:M_cyl}
\end{align}

The transformation of $\mathbb{M}$ under reflections proceeds similarly, with the exception that the pseudovectors $\bm{\Omega}$ and $\bm{\tau}$ do not transform in the same way as the vectors $\bm{V}$ and $\bm{F}$.
Let us consider the reflection $P_z$, which takes $(x, y, z) \to (x, y, -z)$.
In this case, the vectors $\bm{V}$ and $\bm{F}$ transform as expected,
\begin{align}
(F_x, F_y, F_z) \to (F_x, F_y, -F_z)
\end{align}
but pseudovectors $\bm{\Omega}$ and $\bm{\tau}$ transform as
\begin{align}
(\tau_x, \tau_y, \tau_z) \to (-\tau_x, -\tau_y, \tau_z)
\end{align}
since both the torque and angular velocity are defined through a cross product of two vectors (e.g. $\bm{\tau} = \bm{r} \times \bm{F}$).
From this, we follow the steps as in the rotation case to find that the reflected mobility matrix is given by
\begin{align}
\mathbb{M}^{\prime} = 
\begin{bmatrix}
1 & & & & & \\
& 1 & & & & \\
& & -1 & & & \\
& & & -1 & & \\
& & & & -1 & \\
& & & & & 1
\end{bmatrix}
\mathbb{M}
\begin{bmatrix}
1 & & & & & \\
& 1 & & & & \\
& & -1 & & & \\
& & & -1 & & \\
& & & & -1 & \\
& & & & & 1
\end{bmatrix}.
\end{align}
If the fluid/object system is invariant both under $R_z$ and $P_z$, the mobility matrix is constrained to take the form
\begin{align}
\mathbb{M}_{R_z,P_z}=
\begin{bmatrix}
A_{11} & A_{12} & 0 & 0 & 0 & 0 \\
-A_{12} & A_{11} & 0 & 0 & 0 & 0 \\
0 & 0 & A_{33} & 0 & 0 & 0 \\
0 & 0 & 0 & S_{11} & S_{12} & 0 \\
0 & 0 & 0 & -S_{12} & S_{11} & 0 \\
0 & 0 & 0 & 0 & 0 & S_{33}
\end{bmatrix}.
\label{eq:M_cyl_z}
\end{align}

\begin{figure}
    \centering
    \includegraphics[width=0.9\columnwidth]{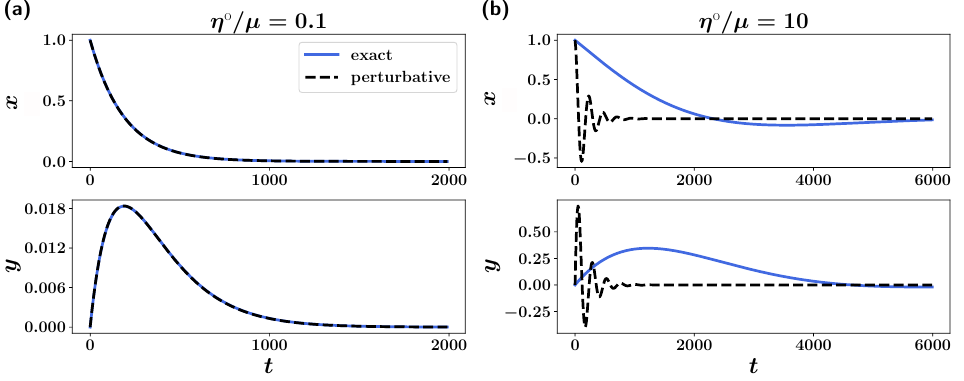}
    \caption{\label{fig:sphere_comp_appendix}
    \textbf{Comparison of sphere trajectory in the perturbative limit to the exact solution.} The perturbative case is from Section~\ref{sec:lift} Fig.~\ref{fig:sphere}b and the exact solution is from Eq.~12 in \cite{everts2024dissipative}. In panel a), the viscosity ratio is taken to be $\eta^{\text{o}}/\mu = 0.1$, and in panel b), $\eta^{\text{o}}/\mu = 10$. The two curves agree well for small odd viscosity (panel a). At high odd viscosity (panel b), the perturbative solution diverges from the exact solution.}
\end{figure}

Note that the parity-violating cylindrical fluid that we consider (see Fig.~\ref{fig:setup}a) is invariant only under these two spatial symmetries. Consequently, $\mathbb{M}_{R_z,P_z}$ must describe the motion of all objects with these same symmetries, for instance, a sphere. 
In fact, we explicitly compute the entries of the mobility matrix of the sphere in Section~\ref{sec:lift} in the presence of $\eta^\text{o}$, and confirm that the mobility matrix takes this form. 
Crucially, this symmetry analysis holds even for high odd viscosity values; the mobility matrix computed for arbitrary odd viscosity in \cite{everts2024dissipative} has the same form. 
We compare our perturbative solution in Fig.~\ref{fig:sphere}b for $\eta^{\text{o}}/\mu = 0.1$ and $\eta^{\text{o}}/\mu = 10$ to the exact solution in Eq.~12 of \cite{everts2024dissipative} in Fig.~\ref{fig:sphere_comp_appendix}. The two curves agree well for small odd viscosity. At higher values of the odd viscosity, the perturbative solution is no longer a good approximation for the sphere trajectory.

Now, if the fluid/object system is not parity-violating -- if it is invariant under reflections across all vertical planes as well (such as under $P_x$) -- the mobility matrix in Eq.~\ref{eq:M_cyl_z} reduces to just a diagonal,
\begin{align}
\mathbb{M}_{R_z, P_z, P_x}=
\begin{bmatrix}
A_{11} & 0 & 0 & 0 & 0 & 0 \\
0 & A_{11} & 0 & 0 & 0 & 0 \\
0 & 0 & A_{33} & 0 & 0 & 0 \\
0 & 0 & 0 & S_{11} & 0 & 0 \\
0 & 0 & 0 & 0 & S_{11} & 0 \\
0 & 0 & 0 & 0 & 0 & S_{33}
\end{bmatrix}.
\label{eq:M_cyl_z_x}
\end{align}
This is the form in Eq.~\ref{eq:M_anisotropic}, which describes both a sphere in an anisotropic fluid and an ellipsoid in an isotropic one.

Interestingly, if the fluid is described only by dissipative viscosities, yet still breaks parity, the mobility matrix in Eq.~\ref{eq:M_cyl_z} is constrained to be symmetric (see Section~\ref{sec:M_symmetry}). In this case, all off-diagonal terms in Eq.~\ref{eq:M_cyl_z} must be zero, and we arrive at the same mobility matrix as in Eq.~\ref{eq:M_cyl_z_x}. That is, in order to get the off-diagonal lift coefficients in Eq.~\ref{eq:M_cyl_z}, the viscosity tensor of our fluid must be both parity-violating and odd.

Let us now turn to the question of chirality. 
We consider a cone with its symmetry axis along the $z$-axis, immersed in an isotropic fluid with only dissipative viscosities. Expanding on Section~\ref{sec:M_parity}, we show that for this fluid/object system, there is a choice of reference point such that the blocks $B$ and $T$ are zero. 
The fluid/object system is invariant under $R_z$ and any reflection across planes containing the $z$-axis. Due to cylindrical symmetry, it is sufficient to require that $\mathbb{M}_{R_z}$ is invariant under $P_x$, which yields the mobility matrix
\begin{align}
\mathbb{M}_{R_z, P_x} = 
\begin{bmatrix}
A_{11} & 0 & 0 & 0 & B_{12} & 0 \\
0 & A_{11} & 0 & -B_{12} & 0 & 0 \\
0 & 0 & A_{33} & 0 & 0 & 0 \\
0 & T_{12} & 0 & S_{11} & 0 & 0 \\
-T_{12} & 0 & 0 & 0 & S_{11} & 0 \\
0 & 0 & 0 & 0 & 0 & S_{33}
\label{eq:M_cyl_x}
\end{bmatrix}.
\end{align}
Additionally, due to the absence of odd viscosities, the mobility matrix must be symmetric:
\begin{align}
\mathbb{M}_{R_z, P_x, \text{even}} = 
\begin{bmatrix}
A_{11} & 0 & 0 & 0 & B_{12} & 0 \\
0 & A_{11} & 0 & -B_{12} & 0 & 0 \\
0 & 0 & A_{33} & 0 & 0 & 0 \\
0 & -B_{12} & 0 & S_{11} & 0 & 0 \\
B_{12} & 0 & 0 & 0 & S_{11} & 0 \\
0 & 0 & 0 & 0 & 0 & S_{33}
\end{bmatrix}.
\end{align}
Finally, by shifting the reference point along the $z$-axis away from the origin to $\bm{r^{*}} = \bm{R} = (0, 0, -B_{12}/S_{11})$, the mobility matrix transforms according to Eq.~\ref{eq:M_refpoint}, and we can remove the terms in blocks $B$ and $T$:
\begin{align}
\mathbb{M}_{R_z, P_x, \text{even}}^{*} = 
\begin{bmatrix}
A_{11} + \frac{3 B_{12}^2}{S_{11}} & 0 & 0 & 0 & 0 & 0 \\
0 & A_{11} + \frac{3 B_{12}^2}{S_{11}} & 0 & 0 & 0 & 0 \\
0 & 0 & A_{33} & 0 & 0 & 0 \\
0 & 0 & 0 & S_{11} & 0 & 0 \\
0 & 0 & 0 & 0 & S_{11} & 0 \\
0 & 0 & 0 & 0 & 0 & S_{33}
\end{bmatrix}.
\end{align}
Note that if the fluid is parity-preserving, but has odd viscosities (Eq.~\ref{eq:M_cyl_x}), the $B$ and $T$ blocks cannot be removed simultaneously by moving the reference point (see Eq.~\ref{eq:M_refpoint}).

Meanwhile, if the cone is immersed in a parity-violating cylindrical fluid, such as the one we consider, the only spatial symmetry obeyed by the fluid/object system is cylindrical symmetry. Hence, the motion of the cone is described by the general mobility matrix in Eq.~\ref{eq:M_cyl}.
In this case, no choice of reference point can cancel out the non-zero $B$ and $T$ blocks. For example, moving the reference point to $\bm{R} = (R_x, R_y, R_z)$ transforms the $T$ block in Eq.~\ref{eq:M_cyl} to
\begin{align}
T_{R_z} \to
\begin{bmatrix}
T_{11} - R_z S_{21} & T_{12} - R_z S_{11} & R_y S_{11} + R_x S_{21} \\
-T_{12} + R_z S_{11} & T_{11} - R_z S_{21} & -R_x S_{11} + R_y S_{21} \\
-R_y S_{33} & R_x S_{33} & T_{33}
\end{bmatrix}.
\end{align}
For generic values of the elements of $T$, there is no choice of $R_z$ that can remove all of the entries. A similar expression holds for $B$. As a result, this cone spins under a force, even though the cone is not chiral, but because the fluid is.

Notice that this effect is distinct from the question of dissipation in the fluid.
If the fluid is described only by even viscosities, the mobility matrix must be symmetric (Sec.~\ref{sec:M_symmetry}), and thus Eq.~\ref{eq:M_cyl} takes the form:
\begin{align}
\mathbb{M}_{R_z, \text{even}}=
\begin{bmatrix}
A_{11} & 0 & 0 & B_{11} & B_{12} & 0 \\
0 & A_{11} & 0 & -B_{12} & B_{11} & 0 \\
0 & 0 & A_{33} & 0 & 0 & B_{33} \\
B_{11} & -B_{12} & 0 & S_{11} & 0 & 0 \\
B_{12} & B_{11} & 0 & 0 & S_{11} & 0 \\
0 & 0 & B_{33} & 0 & 0 & S_{33}
\end{bmatrix}.
\label{eq:M_cyl_even}
\end{align}
Even so, the off diagonal blocks cannot be fully removed (in particular, the diagonals of $B$ and $T$ remain), which leads to rotation under forces even in the absence of odd viscosity.

\section{Numerical values of mobility matrices}
\label{app:M_num}

Below, we provide the numerical values of the symbolic mobility matrices in Sections~\ref{sec:odd_stress}-~\ref{sec:trading}.

The mobility matrix of the triple helix in Eq.~\ref{eq:M_helix} is
\begin{equation}
    \mathbb{M} =
    \begin{bmatrix}
    0.05244 & 0. & 0. & -0.00019 & 0. & 0. \\
    0. & 0.05244 & 0. & 0. & -0.00019 & 0.\\     
    0. & 0. & 0.06464 & 0. & 0. & 0.00612\\
    -0.00019 & 0. & 0. & 0.01062 & 0. & 0.\\     
    0. & -0.00019 & 0. & 0. & 0.01062 & 0.\\     
    0. & 0. & 0.00612 & 0. & 0. & 0.12042
    \end{bmatrix}.
    \label{eq:app_M_helix_num}
\end{equation}

In Section~\ref{sec:spinning}, the triangle object consists of three Stokeslets of radius $a = 0.01$ positioned at the vertices of an equilateral triangle with side-length $b = 1$. For an upright triangle in a standard fluid (Eq.~\ref{eq:M_triangle_up}), the mobility matrix is given by
\begin{equation}
    \mathbb{M} =
    \begin{bmatrix}
  1.80815 & 0. & 0. & 0. & 0. & 0.\\
  0. & 1.79491 & 0. & 0. & 0. & 0.\\
  0. & 0. & 1.80815 & 0. & 0. & 0.\\
  0. & 0. & 0. & 10.53075 & 0. & 0.\\
  0. & 0. & 0. & 0. & 5.28527 & 0.\\
  0. & 0. & 0. & 0. & 0. & 10.53075\\
    \end{bmatrix}.
\end{equation}

The mobility matrix of the triangle changes if it is immersed in a fluid with odd viscosity. For $\eta^{\text{o}}/\mu = 0.1$, 
\begin{equation}
    \mathbb{M} =
    \begin{bmatrix}
      1.80815 & 0.00897 & 0. & 0.00006 & 0. & 0.\\
      -0.00897 & 1.79491 & 0. & 0. & 0. & 0.\\
      0. & 0. & 1.80815 & 0. & 0. & -0.00006\\
      -0.00006 & 0. & -0. & 10.53088 & 0.02672 & 0.\\
      0. & 0. & 0. & -0.02672 & 5.28527 & 0.\\
      0. & 0. & 0.00006 & 0. & 0. & 10.53101\\
    \end{bmatrix}.
\end{equation}

The mobility matrix for the conical helix in Section~\ref{sec:trading} is
\begin{equation}
    \mathbb{M} =
    \begin{bmatrix}
      0.11767 & 0. & 0. & 0.00009 & -0.02061 & 0.\\
      0. & 0.11767 & 0. & 0.02061 & 0.00009 & 0.\\
      0. & 0. & 0.0378 & 0. & 0. & -0.0002\\
      0.00009 & 0.02061 & 0. & 0.00536 & 0. & 0.\\
      -0.02061 & 0.00009 & 0. & 0. & 0.00536 & 0.\\
      0. & 0. & -0.0002 & 0. & 0. & 0.00614\\
    \end{bmatrix}.
\label{eq:app_M_con_helix_num}
\end{equation}
The mobility matrix for the pushpin-like object in Section~\ref{sec:trading} is
\begin{equation}    
\mathbb{M} = 
    \begin{bmatrix}
  0.41715 & 0. & 0. & 0.00008 & -0.00624 & 0.\\
  0. & 0.41715 & 0. & 0.00624 & 0.00008 & 0.\\
  0. & 0. & 0.41217 & 0. & 0. & -0.00025\\
  0.00008 & 0.00624 & 0. & 0.32767 & 0. & 0.\\
  -0.00624 & 0.00008 & 0. & 0. & 0.32767 & 0.\\
  0. & 0. & -0.00025 & 0. & 0. & 0.5543\\
\end{bmatrix}.
\label{eq:app_M_pushpin_num}
\end{equation}
Both of these have the same form as Eq.~\ref{eq:M_cyl_even}, as desired.

\bibliographystyle{jfm}

\begin{thebibliography}{62}
\expandafter\ifx\csname natexlab\endcsname\relax\def\natexlab#1{#1}\fi
\def\au#1{#1} \def\ed#1{#1} \def\yr#1{#1}\def\at#1{#1}\def\jt#1{\textit{#1}}
  \def\bt#1{#1}\def\bvol#1{\textbf{#1}} \def\vol#1{#1} \def\pg#1{#1}
  \def\publ#1{#1}\def\arxiv#1{#1}\def\org#1{#1}\def\st#1{\textit{#1}}

\bibitem[Avron(1998)]{avron1998odd}
{\sc \au{Avron, JE}} \yr{1998}  \at{Odd viscosity}.  \jt{Journal of statistical
  physics}  \bvol{92},  \pg{543--557}.

\bibitem[Bandurin {\em et~al.\/}(2016)Bandurin, Torre, Kumar, Shalom, Tomadin,
  Principi, Auton, Khestanova, Novoselov, Grigorieva, Ponomarenko, Geim \&
  Polini]{Bandurin2016}
{\sc \au{Bandurin, D.~A.}, \au{Torre, I.}, \au{Kumar, R.~K.}, \au{Shalom,
  M.~Ben}, \au{Tomadin, A.}, \au{Principi, A.}, \au{Auton, G.~H.},
  \au{Khestanova, E.}, \au{Novoselov, K.~S.}, \au{Grigorieva, I.~V.},
  \au{Ponomarenko, L.~A.}, \au{Geim, A.~K.} \& \au{Polini, M.}} \yr{2016}
  \at{Negative local resistance caused by viscous electron backflow in
  graphene}.  \jt{Science}  \bvol{351}~(6277),  \pg{1055--1058}.

\bibitem[Banerjee {\em et~al.\/}(2017)Banerjee, Souslov, Abanov \&
  Vitelli]{banerjee2017odd}
{\sc \au{Banerjee, D.}, \au{Souslov, A.}, \au{Abanov, A.~G.} \& \au{Vitelli,
  V.}} \yr{2017}  \at{Odd viscosity in chiral active fluids}.  \jt{Nature
  communications}  \bvol{8}~(1),  \pg{1--12}.

\bibitem[Berdyugin {\em et~al.\/}(2019)Berdyugin, Xu, Pellegrino,
  Krishna~Kumar, Principi, Torre, Ben~Shalom, Taniguchi, Watanabe, Grigorieva,
  Polini, Geim \& Bandurin]{Berdyugin2019measuring}
{\sc \au{Berdyugin, A.~I.}, \au{Xu, S.~G.}, \au{Pellegrino, F. M.~D.},
  \au{Krishna~Kumar, R.}, \au{Principi, A.}, \au{Torre, I.}, \au{Ben~Shalom,
  M.}, \au{Taniguchi, T.}, \au{Watanabe, K.}, \au{Grigorieva, I.~V.},
  \au{Polini, M.}, \au{Geim, A.~K.} \& \au{Bandurin, D.~A.}} \yr{2019}
  \at{Measuring {H}all viscosity of graphene{\textquoteright}s electron fluid}.
   \jt{Science}  \bvol{364}~(6436),  \pg{162--165}.

\bibitem[Bloomfield {\em et~al.\/}(1967)Bloomfield, Dalton \&
  Van~Holde]{bloomfield1967frictional}
{\sc \au{Bloomfield, V}, \au{Dalton, WO} \& \au{Van~Holde, KE}} \yr{1967}
  \at{Frictional coefficients of multisubunit structures. {I}. {T}heory}.
  \jt{Biopolymers: Original Research on Biomolecules}  \bvol{5}~(2),
  \pg{135--148}.

\bibitem[Brenner(1965)]{brenner1965coupling}
{\sc \au{Brenner, Howard}} \yr{1965}  \at{Coupling between the translational
  and rotational brownian motions of rigid particles of arbitrary shape {I}.
  {H}elicoidally isotropic particles}.  \jt{Journal of colloid science}
  \bvol{20}~(2),  \pg{104--122}.

\bibitem[Burgers(1938)]{burgers1938second}
{\sc \au{Burgers, J.~M.}} \yr{1938}  \bt{ \at{Second report in viscosity and
  plasticity of the {A}msterdam {A}cademy of {S}ciences}}. chap. III.
  \publ{Nordemann Publishing Company}.

\bibitem[Chajwa {\em et~al.\/}(2019)Chajwa, Menon \&
  Ramaswamy]{Chajwa2019kepler}
{\sc \au{Chajwa, Rahul}, \au{Menon, Narayanan} \& \au{Ramaswamy, Sriram}}
  \yr{2019}  \at{Kepler orbits in pairs of disks settling in a viscous fluid}.
  \jt{Phys. Rev. Lett.}  \bvol{122},  \pg{224501}.

\bibitem[Chapman(1939)]{Chapman1939}
{\sc \au{Chapman, Sydney}} \yr{1939} {\em The mathematical theory of
  non-uniform gases; an account of the kinetic theory of viscosity, thermal
  conduction, and diffusion in gases\/}.  \publ{Cambridge Eng.: University
  Press}.

\bibitem[Condiff \& Dahler(1964)]{Condiff1964}
{\sc \au{Condiff, Duane~W.} \& \au{Dahler, John~S.}} \yr{1964}  \at{Fluid
  mechanical aspects of antisymmetric stress}.  \jt{Physics of Fluids}
  \bvol{7}~(6),  \pg{842}.

\bibitem[Everts \& Cichocki(2024)]{everts2024dissipative}
{\sc \au{Everts, Jeffrey~C} \& \au{Cichocki, Bogdan}} \yr{2024}
  \at{Dissipative effects in odd viscous stokes flow around a single sphere}.
  \jt{Physical Review Letters}  \bvol{132}~(21),  \pg{218303}.

\bibitem[Fruchart {\em et~al.\/}(2023)Fruchart, Scheibner \&
  Vitelli]{fruchart2023odd}
{\sc \au{Fruchart, Michel}, \au{Scheibner, Colin} \& \au{Vitelli, Vincenzo}}
  \yr{2023}  \at{Odd viscosity and odd elasticity}.  \jt{Annual Review of
  Condensed Matter Physics}  \bvol{14},  \pg{471--510}.

\bibitem[Ganeshan \& Abanov(2017)]{ganeshan2017odd}
{\sc \au{Ganeshan, Sriram} \& \au{Abanov, Alexander~G}} \yr{2017}  \at{Odd
  viscosity in two-dimensional incompressible fluids}.  \jt{Physical review
  fluids}  \bvol{2}~(9),  \pg{094101}.

\bibitem[Goldfriend {\em et~al.\/}(2017)Goldfriend, Diamant \&
  Witten]{Goldfriend2017screening}
{\sc \au{Goldfriend, Tomer}, \au{Diamant, Haim} \& \au{Witten, Thomas~A.}}
  \yr{2017}  \at{Screening, hyperuniformity, and instability in the
  sedimentation of irregular objects}.  \jt{Phys. Rev. Lett.}  \bvol{118},
  \pg{158005}.

\bibitem[Gonz{\'a}lez(2010)]{gonzalez2010measurement}
{\sc \au{Gonz{\'a}lez, {\'A}lvaro}} \yr{2010}  \at{Measurement of areas on a
  sphere using {F}ibonacci and latitude--longitude lattices}.  \jt{Mathematical
  Geosciences}  \bvol{42},  \pg{49--64}.

\bibitem[Guazzelli \& Morris(2009)]{Guazzelli2009}
{\sc \au{Guazzelli, Elisabeth} \& \au{Morris, Jeffrey~F.}} \yr{2009} {\em A
  Physical Introduction to Suspension Dynamics\/}.  \publ{Cambridge University
  Press}.

\bibitem[Guyon {\em et~al.\/}(2015)Guyon, Hulin, Petit \&
  Mitescu]{guyon2015physical}
{\sc \au{Guyon, Etienne}, \au{Hulin, Jean~Pierre}, \au{Petit, Luc} \&
  \au{Mitescu, Catalin~D}} \yr{2015} {\em Physical hydrodynamics\/}.
  \publ{Oxford university press}.

\bibitem[Han {\em et~al.\/}(2021)Han, Fruchart, Scheibner, Vaikuntanathan,
  De~Pablo \& Vitelli]{han2021fluctuating}
{\sc \au{Han, Ming}, \au{Fruchart, Michel}, \au{Scheibner, Colin},
  \au{Vaikuntanathan, Suriyanarayanan}, \au{De~Pablo, Juan~J} \& \au{Vitelli,
  Vincenzo}} \yr{2021}  \at{Fluctuating hydrodynamics of chiral active fluids}.
   \jt{Nature Physics}  \bvol{17}~(11),  \pg{1260--1269}.

\bibitem[Happel(1983)]{Happel1983low}
{\sc \au{Happel, John}} \yr{1983} {\em Low Reynolds number hydrodynamics with
  special applications to particulate media\/}, 1st edn.  \publ{The Hague;
  Boston: Hingham, MA, USA: M. Nijhoff; Distributed by Kluwer Boston}.

\bibitem[Hargus {\em et~al.\/}(2020)Hargus, Klymko, Epstein \&
  Mandadapu]{Hargus2020}
{\sc \au{Hargus, Cory}, \au{Klymko, Katherine}, \au{Epstein, Jeffrey~M.} \&
  \au{Mandadapu, Kranthi~K.}} \yr{2020}  \at{Time reversal symmetry breaking
  and odd viscosity in active fluids: Green–{K}ubo and {NEMD} results}.
  \jt{The Journal of Chemical Physics}  \bvol{152}~(20).

\bibitem[Herron {\em et~al.\/}(1975)Herron, Davis \&
  Bretherton]{herron1975sedimentation}
{\sc \au{Herron, Isom~H}, \au{Davis, Stephen~H} \& \au{Bretherton, Francis~P}}
  \yr{1975}  \at{On the sedimentation of a sphere in a centrifuge}.
  \jt{Journal of Fluid Mechanics}  \bvol{68}~(2),  \pg{209--234}.

\bibitem[Hosaka {\em et~al.\/}(2023)Hosaka, Golestanian \&
  Vilfan]{hosaka2023lorentz}
{\sc \au{Hosaka, Yuto}, \au{Golestanian, Ramin} \& \au{Vilfan, Andrej}}
  \yr{2023}  \at{Lorentz reciprocal theorem in fluids with odd viscosity}.
  \jt{Phys. Rev. Lett.}  \bvol{131},  \pg{178303}.

\bibitem[Khain {\em et~al.\/}(2023)Khain, Fruchart \&
  Vitelli]{khain2023viscous}
{\sc \au{Khain, Tali}, \au{Fruchart, Michel} \& \au{Vitelli, Vincenzo}}
  \yr{2023}  \at{Viscous tweezers: controlling particles with viscosity}.
  \jt{arXiv preprint arXiv:2307.04948} .

\bibitem[Khain {\em et~al.\/}(2022)Khain, Scheibner, Fruchart \&
  Vitelli]{khain2022stokes}
{\sc \au{Khain, Tali}, \au{Scheibner, Colin}, \au{Fruchart, Michel} \&
  \au{Vitelli, Vincenzo}} \yr{2022}  \at{Stokes flows in three-dimensional
  fluids with odd and parity-violating viscosities}.  \jt{Journal of Fluid
  Mechanics}  \bvol{934}.

\bibitem[Kim \& Karrila(1991)]{KimKarrila}
{\sc \au{Kim, Sangtae} \& \au{Karrila, Seppo~J.}} \yr{1991} {\em
  Microhydrodynamics\/}.  \publ{Butterworth-Heinemann}.

\bibitem[Kirkwood \& Riseman(1948)]{kirkwood1948intrinsic}
{\sc \au{Kirkwood, John~G} \& \au{Riseman, Jacob}} \yr{1948}  \at{The intrinsic
  viscosities and diffusion constants of flexible macromolecules in solution}.
  \jt{The Journal of Chemical Physics}  \bvol{16}~(6),  \pg{565--573}.

\bibitem[Korving {\em et~al.\/}(1967)Korving, Hulsman, Scoles, Knaap \&
  Beenakker]{Korving1967influence}
{\sc \au{Korving, J.}, \au{Hulsman, H.}, \au{Scoles, G.}, \au{Knaap, H.F.P.} \&
  \au{Beenakker, J.J.M.}} \yr{1967}  \at{The influence of a magnetic field on
  the transport properties of gases of polyatomic molecules;: Part {I},
  {V}iscosity}.  \jt{Physica}  \bvol{36}~(2),  \pg{177 -- 197}.

\bibitem[Krapf {\em et~al.\/}(2009)Krapf, Witten \& Keim]{krapf2009chiral}
{\sc \au{Krapf, Nathan~W}, \au{Witten, Thomas~A} \& \au{Keim, Nathan~C}}
  \yr{2009}  \at{Chiral sedimentation of extended objects in viscous media}.
  \jt{Physical Review E}  \bvol{79}~(5),  \pg{056307}.

\bibitem[Lapa \& Hughes(2014)]{Lapa2014swimming}
{\sc \au{Lapa, Matthew~F.} \& \au{Hughes, Taylor~L.}} \yr{2014}  \at{Swimming
  at low {R}eynolds number in fluids with odd, or {H}all, viscosity}.
  \jt{Phys. Rev. E}  \bvol{89},  \pg{043019}.

\bibitem[Lauga(2016)]{lauga2016bacterial}
{\sc \au{Lauga, Eric}} \yr{2016}  \at{Bacterial hydrodynamics}.  \jt{Annual
  Review of Fluid Mechanics}  \bvol{48},  \pg{105--130}.

\bibitem[Lauga(2020)]{lauga2020fluid}
{\sc \au{Lauga, Eric}} \yr{2020} {\em The fluid dynamics of cell motility\/}.
  \publ{Cambridge University Press}.

\bibitem[Lauga \& Powers(2009)]{lauga2009hydrodynamics}
{\sc \au{Lauga, Eric} \& \au{Powers, Thomas~R}} \yr{2009}  \at{The
  hydrodynamics of swimming microorganisms}.  \jt{Reports on progress in
  physics}  \bvol{72}~(9),  \pg{096601}.

\bibitem[Levine \& Lubensky(2001)]{levine2001response}
{\sc \au{Levine, Alex~J} \& \au{Lubensky, TC}} \yr{2001}  \at{Response function
  of a sphere in a viscoelastic two-fluid medium}.  \jt{Physical Review E}
  \bvol{63}~(4),  \pg{041510}.

\bibitem[Lier {\em et~al.\/}(2023)Lier, Duclut, Bo, Armas, J{\"u}licher \&
  Sur{\'o}wka]{lier2023lift}
{\sc \au{Lier, Ruben}, \au{Duclut, Charlie}, \au{Bo, Stefano}, \au{Armas, Jay},
  \au{J{\"u}licher, Frank} \& \au{Sur{\'o}wka, Piotr}} \yr{2023}  \at{Lift
  force in odd compressible fluids}.  \jt{Physical Review E}  \bvol{108}~(2),
  \pg{L023101}.

\bibitem[Makino \& Doi(2003)]{makino2003sedimentation}
{\sc \au{Makino, Masato} \& \au{Doi, Masao}} \yr{2003}  \at{Sedimentation of a
  particle with translation--rotation coupling}.  \jt{Journal of the Physical
  Society of Japan}  \bvol{72}~(11),  \pg{2699--2701}.

\bibitem[Markovich \& Lubensky(2021)]{markovich2021odd}
{\sc \au{Markovich, Tomer} \& \au{Lubensky, Tom~C}} \yr{2021}  \at{Odd
  viscosity in active matter: microscopic origin and 3d effects}.  \jt{Physical
  Review Letters}  \bvol{127}~(4),  \pg{048001}.

\bibitem[Markovich \& Lubensky(2022)]{markovich2022non}
{\sc \au{Markovich, Tomer} \& \au{Lubensky, Tom~C}} \yr{2022}  \at{Non
  reciprocal odd viscosity: Coarse graining the kinetic energy and exceptional
  instability}.  \jt{arXiv preprint arXiv:2211.06901} .

\bibitem[Masoud \& Stone(2019)]{masoud2019reciprocal}
{\sc \au{Masoud, Hassan} \& \au{Stone, Howard~A}} \yr{2019}  \at{The reciprocal
  theorem in fluid dynamics and transport phenomena}.  \jt{Journal of Fluid
  Mechanics}  \bvol{879},  \pg{P1}.

\bibitem[Meakin \& Deutch(1987)]{meakin1987properties}
{\sc \au{Meakin, Paul} \& \au{Deutch, John~M}} \yr{1987}  \at{Properties of the
  fractal measure describing the hydrodynamic force distributions for fractal
  aggregates moving in a quiescent fluid}.  \jt{The Journal of chemical
  physics}  \bvol{86}~(8),  \pg{4648--4656}.

\bibitem[Miara {\em et~al.\/}(2024)Miara, Vaquero-Stainer, Pihler-Puzovi{\'c},
  Heil \& Juel]{miara2024dynamics}
{\sc \au{Miara, Tymoteusz}, \au{Vaquero-Stainer, Christian},
  \au{Pihler-Puzovi{\'c}, Draga}, \au{Heil, Matthias} \& \au{Juel, Anne}}
  \yr{2024}  \at{Dynamics of inertialess sedimentation of a rigid u-shaped
  disk}.  \jt{Communications Physics}  \bvol{7}~(1),  \pg{47}.

\bibitem[Mittal \& Iaccarino(2005)]{mittal2005immersed}
{\sc \au{Mittal, Rajat} \& \au{Iaccarino, Gianluca}} \yr{2005}  \at{Immersed
  boundary methods}.  \jt{Annu. Rev. Fluid Mech.}  \bvol{37},  \pg{239--261}.

\bibitem[Morozov {\em et~al.\/}(2017)Morozov, Mirzae, Kenneth \&
  Leshansky]{morozov2017dynamics}
{\sc \au{Morozov, Konstantin~I}, \au{Mirzae, Yoni}, \au{Kenneth, Oded} \&
  \au{Leshansky, Alexander~M}} \yr{2017}  \at{Dynamics of arbitrary shaped
  propellers driven by a rotating magnetic field}.  \jt{Physical Review Fluids}
   \bvol{2}~(4),  \pg{044202}.

\bibitem[Mowitz \& Witten(2017)]{mowitz2017predicting}
{\sc \au{Mowitz, Aaron~J} \& \au{Witten, TA}} \yr{2017}  \at{Predicting
  tensorial electrophoretic effects in asymmetric colloids}.  \jt{Physical
  Review E}  \bvol{96}~(6),  \pg{062613}.

\bibitem[Nakagawa(1956)]{Yoshinari1956Kinetic}
{\sc \au{Nakagawa, Yoshinari}} \yr{1956}  \at{The kinetic theory of gases for
  the rotating system.}  \jt{Journal of Physics of the Earth}  \bvol{4}~(3),
  \pg{105--111}.

\bibitem[Palusa {\em et~al.\/}(2018)Palusa, De~Graaf, Brown \&
  Morozov]{palusa2018sedimentation}
{\sc \au{Palusa, Martina}, \au{De~Graaf, Joost}, \au{Brown, Aidan} \&
  \au{Morozov, Alexander}} \yr{2018}  \at{Sedimentation of a rigid helix in
  viscous media}.  \jt{Physical Review Fluids}  \bvol{3}~(12),  \pg{124301}.

\bibitem[Park \& Park(2021)]{park2021thermodynamic}
{\sc \au{Park, Jong-Min} \& \au{Park, Hyunggyu}} \yr{2021}  \at{Thermodynamic
  uncertainty relation in the overdamped limit with a magnetic {L}orentz
  force}.  \jt{Physical Review Research}  \bvol{3}~(4),  \pg{043005}.

\bibitem[Pozrikidis(1992)]{pozrikidis1992boundary}
{\sc \au{Pozrikidis, Constantine}} \yr{1992} {\em Boundary integral and
  singularity methods for linearized viscous flow\/}.  \publ{Cambridge
  University Press}.

\bibitem[Purcell(1977)]{Purcell1977life}
{\sc \au{Purcell, E.~M.}} \yr{1977}  \at{Life at low {R}eynolds number}.
  \jt{American Journal of Physics}  \bvol{45}~(1),  \pg{3--11}.

\bibitem[Ramaswamy(2001)]{Ramaswamy2001}
{\sc \au{Ramaswamy, Sriram}} \yr{2001}  \at{Issues in the statistical mechanics
  of steady sedimentation}.  \jt{Advances in Physics}  \bvol{50}~(3),
  \pg{297--341}.

\bibitem[Reeves {\em et~al.\/}(2021)Reeves, Aranson \&
  Vlahovska]{reeves2021emergence}
{\sc \au{Reeves, Cody~J}, \au{Aranson, Igor~S} \& \au{Vlahovska, Petia~M}}
  \yr{2021}  \at{Emergence of lanes and turbulent-like motion in active spinner
  fluid}.  \jt{Communications Physics}  \bvol{4}~(1),  \pg{92}.

\bibitem[Reynolds {\em et~al.\/}(2023)Reynolds, Monteiro \&
  Ganeshan]{Reynolds2023}
{\sc \au{Reynolds, Dylan}, \au{Monteiro, Gustavo~M.} \& \au{Ganeshan, Sriram}}
  \yr{2023} Three dimensional odd viscosity in ferrofluids with
  vorticity-magnetization coupling,  \arxiv{arXiv: 2301.07096}.

\bibitem[Rotne \& Prager(1969)]{rotne1969variational}
{\sc \au{Rotne, Jens} \& \au{Prager, Stephen}} \yr{1969}  \at{Variational
  treatment of hydrodynamic interaction in polymers}.  \jt{The Journal of
  Chemical Physics}  \bvol{50}~(11),  \pg{4831--4837}.

\bibitem[Soni {\em et~al.\/}(2019)Soni, Bililign, Magkiriadou, Sacanna,
  Bartolo, Shelley \& Irvine]{soni2019odd}
{\sc \au{Soni, V.}, \au{Bililign, E.~S.}, \au{Magkiriadou, S.}, \au{Sacanna,
  S.}, \au{Bartolo, D.}, \au{Shelley, M.~J.} \& \au{Irvine, W. T.~M.}}
  \yr{2019}  \at{The odd free surface flows of a colloidal chiral fluid}.
  \jt{Nature Physics}  \bvol{15}~(11),  \pg{1188--1194}.

\bibitem[Tanzosh \& Stone(1994)]{tanzosh1994motion}
{\sc \au{Tanzosh, John~P} \& \au{Stone, Howard~A}} \yr{1994}  \at{Motion of a
  rigid particle in a rotating viscous flow: an integral equation approach}.
  \jt{Journal of Fluid Mechanics}  \bvol{275},  \pg{225--256}.

\bibitem[Tanzosh \& Stone(1995)]{tanzosh1995transverse}
{\sc \au{Tanzosh, John~P} \& \au{Stone, Howard~A}} \yr{1995}  \at{Transverse
  motion of a disk through a rotating viscous fluid}.  \jt{Journal of Fluid
  Mechanics}  \bvol{301},  \pg{295--324}.

\bibitem[Taylor(1951)]{Taylor1951}
{\sc \au{Taylor, Geoffrey~Ingram}} \yr{1951}  \at{Analysis of the swimming of
  microscopic organisms}.  \jt{Proceedings of the Royal Society of London.
  Series A. Mathematical and Physical Sciences}  \bvol{209}~(1099),
  \pg{447--461}.

\bibitem[Tsai {\em et~al.\/}(2005)Tsai, Ye, Rodriguez, Gollub \&
  Lubensky]{Tsai2005chiral}
{\sc \au{Tsai, J.-C.}, \au{Ye, Fangfu}, \au{Rodriguez, Juan}, \au{Gollub,
  J.~P.} \& \au{Lubensky, T.~C.}} \yr{2005}  \at{A chiral granular gas}.
  \jt{Phys. Rev. Lett.}  \bvol{94},  \pg{214301}.

\bibitem[Verzicco(2023)]{Verzicco2023}
{\sc \au{Verzicco, Roberto}} \yr{2023}  \at{Immersed boundary methods:
  Historical perspective and future outlook}.  \jt{Annual Review of Fluid
  Mechanics}  \bvol{55}~(1),  \pg{129–155}.

\bibitem[Wiegmann \& Abanov(2014)]{wiegmann2014anomalous}
{\sc \au{Wiegmann, P.} \& \au{Abanov, A.~G.}} \yr{2014}  \at{Anomalous
  hydrodynamics of two-dimensional vortex fluids}.  \jt{Physical review
  letters}  \bvol{113}~(3),  \pg{034501}.

\bibitem[Witten \& Diamant(2020)]{witten2020review}
{\sc \au{Witten, Thomas~A} \& \au{Diamant, Haim}} \yr{2020}  \at{A review of
  shaped colloidal particles in fluids: anisotropy and chirality}.  \jt{Reports
  on Progress in Physics}  \bvol{83}~(11),  \pg{116601}.

\bibitem[Yamakawa(1970)]{yamakawa1970transport}
{\sc \au{Yamakawa, Hiromi}} \yr{1970}  \at{Transport properties of polymer
  chains in dilute solution: hydrodynamic interaction}.  \jt{The Journal of
  Chemical Physics}  \bvol{53}~(1),  \pg{436--443}.

\bibitem[Yuan \& Olvera de~la Cruz(2023)]{yuan2023stokesian}
{\sc \au{Yuan, Hang} \& \au{Olvera de~la Cruz, Monica}} \yr{2023}
  \at{Stokesian dynamics with odd viscosity}.  \jt{Physical Review Fluids}
  \bvol{8}~(5),  \pg{054101}.

\end{thebibliography}

\providecommand{\noopsort}[1]{}\providecommand{\singleletter}[1]{#1}%

\end{document}